# Upper Limits on Gravitational Waves from Scorpius X-1 from a Model-based Cross-correlation Search in Advanced LIGO Data

B. P. Abbott[1], R. Abbott[1], T. D. Abbott[2], F. Acernese[3,4], K. Ackley[5], C. Adams[6], T. Adams[7], P. Addesso[8], R. X. Adhikari[1], V. B. Adya[9], C. Affeldt[9], M. Afrough[10], B. Agarwal[11], M. Agathos[12], K. Agatsuma[13], N. Aggarwal[14], O. D. Aguiar[15], L. Aiello[16,17], A. Ain[18], P. Ajith[19], B. Allen[9,20,21], G. Allen[11], A. Allocca[22,23], P. A. Altin[24], A. Amato[25], A. Ananyeva[1], S. B. Anderson[1], W. G. Anderson[20], S. Antier[26], S. Appert[1], K. Arai[1], M. C. Araya[1], J. S. Areeda[27], N. Arnaud[26,28], K. G. Arun[29], S. Ascenzi[17,30], G. Ashton[9], M. Ast[31], S. M. Aston[6], P. Astone[32], P. Aufmuth[21], C. Aulbert[9], K. AultONeal[33], A. Avila-Alvarez[27], S. Babak[34], P. Bacon[35], M. K. M. Bader[13], S. Bae[36], P. T. Baker[37,38], F. Baldaccini[39,40], G. Ballardin[28], S. W. Ballmer[41], S. Banagiri[42], J. C. Barayoga[1], S. E. Barclay[43], B. C. Barish[1], D. Barker[44], F. Barone[3,4], B. Barr[43], L. Barsotti[14], M. Barsuglia[35], D. Barta[45], J. Bartlett[44], I. Bartos[46], R. Bassiri[47], A. Basti[22,23], J. C. Batch[44], C. Baune[9], M. Bawaj[48,40], M. Bazzan[49,50], B. Bécsy[51], C. Beer[9], M. Bejger[52], I. Belahcene[26], A. S. Bell[43], B. K. Berger[1], G. Bergmann[9], C. P. L. Berry[53], D. Bersanetti[54,55], A. Bertolini[13], J. Betzwieser[6], S. Bhagwat[41], R. Bhandare[56], I. A. Bilenko[57], G. Billingsley[1], C. R. Billman[5], J. Birch[6], R. Birney[58], O. Birnholtz[9], S. Biscans[14], A. Bisht[21], M. Bitossi[23,28], C. Biwer[41], M. A. Bizouard[26], J. K. Blackburn[1], J. Blackman[59], C. D. Blair[60], D. G. Blair[60], R. M. Blair[44], S. Bloemen[61], O. Bock[9], N. Bode[9], M. Boer[62], G. Bogaert[62], A. Bohe[34], F. Bondu[63], R. Bonnand[7], B. A. Boom[13], R. Bork[1], V. Boschi[22,23], S. Bose[18,64], Y. Bouffanais[35], A. Bozzi[28], C. Bradaschia[23], P. R. Brady[20], V. B. Braginsky[57,149], M. Branchesi[65,66], J. E. Brau[67], T. Briant[68], A. Brillet[62], M. Brinkmann[9], V. Brisson[26], P. Brockill[20], J. E. Broida[69], A. F. Brooks[1], D. A. Brown[41], D. D. Brown[53], N. M. Brown[14], S. Brunett[1], C. C. Buchanan[2], A. Buikema[14], T. Bulik[70], H. J. Bulten[13,71], A. Buonanno[34,72], D. Buskulic[7], C. Buy[35], R. L. Byer[47], M. Cabero[9], L. Cadonati[73], G. Cagnoli[25,74], C. Cahillane[1], J. Calderón Bustillo[73], T. A. Callister[1], E. Calloni[4,75], J. B. Camp[76], M. Canepa[54,55], P. Canizares[61], K. C. Cannon[77], H. Cao[78], J. Cao[79], C. D. Capano[9], E. Capocasa[35], F. Carbognani[28], S. Caride[80], M. F. Carney[81], J. Casanueva Diaz[26], C. Casentini[17,30], S. Caudill[20], M. Cavaglià[10], F. Cavalier[26], R. Cavalieri[28], G. Cella[23], C. B. Cepeda[1], L. Cerboni Baiardi[65,66], G. Cerretani[22,23], E. Cesarini[17,30], S. J. Chamberlin[82], M. Chan[43], S. Chao[83], P. Charlton[84], E. Chassande-Mottin[35], D. Chatterjee[20], K. Chatziioannou[85], B. D. Cheeseboro[37,38], H. Y. Chen[86], Y. Chen[59], H.-P. Cheng[5], A. Chincarini[55], A. Chiummo[28], T. Chmiel[81], H. S. Cho[87], M. Cho[72], J. H. Chow[24], N. Christensen[62,69], Q. Chu[60], A. J. K. Chua[12], S. Chua[68], A. K. W. Chung[88], S. Chung[60], G. Ciani[5], R. Ciolfi[89,90], C. E. Cirelli[47], A. Cirone[54,55], F. Clara[44], J. A. Clark[73], F. Cleva[62], C. Cocchieri[10], E. Coccia[16,17], P.-F. Cohadon[68], A. Colla[32,91], C. G. Collette[92], L. R. Cominsky[93], M. Constancio Jr.[15], L. Conti[50], S. J. Cooper[53], P. Corban[6], T. R. Corbitt[2], K. R. Corley[46], N. Cornish[94], A. Corsi[80], S. Cortese[28], C. A. Costa[15], M. W. Coughlin[69], S. B. Coughlin[95,96], J.-P. Coulon[62], S. T. Countryman[46], P. Couvares[1], P. B. Covas[97], E. E. Cowan[73], D. M. Coward[60], M. J. Cowart[6], D. C. Coyne[1], R. Coyne[80], J. D. E. Creighton[20], T. D. Creighton[98], J. Cripe[2], S. G. Crowder[99], T. J. Cullen[27], A. Cumming[43], L. Cunningham[43], E. Cuoco[28], T. Dal Canton[76], S. L. Danilishin[9,21], S. D'Antonio[17], K. Danzmann[9,21], A. Dasgupta[100], C. F. Da Silva Costa[5], V. Dattilo[28], I. Dave[56], M. Davier[26], D. Davis[41], E. J. Daw[101], B. Day[73], S. De[41], D. DeBra[47], E. Deelman[102], J. Degallaix[25], M. De Laurentis[4,75], S. Deléglise[68], W. Del Pozzo[22,23,53], T. Denker[9], T. Dent[9], V. Dergachev[34], R. De Rosa[4,75], R. T. DeRosa[6], R. DeSalvo[103], J. Devenson[58], R. C. Devine[37,38], S. Dhurandhar[18], M. C. Díaz[98], L. Di Fiore[4], M. Di Giovanni[90,104], T. Di Girolamo[4,46,75], A. Di Lieto[22,23], S. Di Pace[32,91], I. Di Palma[32,91], F. Di Renzo[22,23], Z. Doctor[86], V. Dolique[25], F. Donovan[14], K. L. Dooley[10], S. Doravari[9], I. Dorrington[96], R. Douglas[43], M. Dovale Álvarez[53], T. P. Downes[20], M. Drago[9], R. W. P. Drever[1,150], J. C. Driggers[44], Z. Du[79], M. Ducrot[7], J. Duncan[95], S. E. Dwyer[44], T. B. Edo[101], M. C. Edwards[69], A. Effler[6], H.-B. Eggenstein[9], P. Ehrens[1], J. Eichholz[1], S. S. Eikenberry[5], R. A. Eisenstein[14], R. C. Essick[14], Z. B. Etienne[37,38], T. Etzel[1], M. Evans[14], T. M. Evans[6], M. Factourovich[46], V. Fafone[16,17,30], H. Fair[41], S. Fairhurst[96], X. Fan[79], S. Farinon[55], B. Farr[86], W. M. Farr[53], E. J. Fauchon-Jones[96], M. Favata[105], M. Fays[96], H. Fehrmann[9], J. Feicht[1], M. M. Fejer[47], A. Fernandez-Galiana[14], I. Ferrante[22,23], E. C. Ferreira[15], F. Ferrini[28], F. Fidecaro[22,23], I. Fiori[28], D. Fiorucci[35], R. P. Fisher[41], R. Flaminio[25,106], M. Fletcher[43], H. Fong[85], P. W. F. Forsyth[24], S. S. Forsyth[73], J.-D. Fournier[62], S. Frasca[32,91], F. Frasconi[23], Z. Frei[51], A. Freise[53], R. Frey[67], V. Frey[26], E. M. Fries[1], P. Fritschel[14], V. V. Frolov[6], P. Fulda[5,76], M. Fyffe[6], H. Gabbard[9], M. Gabel[107], B. U. Gadre[18], S. M. Gaebel[53], J. R. Gair[108], D. K. Galloway[109], L. Gammaitoni[39], M. R. Ganija[78], S. G. Gaonkar[18], F. Garufi[4,75], S. Gaudio[33], G. Gaur[110], V. Gayathri[111], N. Gehrels[76,151], G. Gemme[55], E. Genin[28], A. Gennai[23], D. George[11], J. George[56], L. Gergely[112], V. Germain[7], S. Ghonge[73], Abhirup Ghosh[19], Archisman Ghosh[13,19], S. Ghosh[13,61], J. A. Giaime[2,6], K. D. Giardina[6], A. Giazotto[23], K. Gill[33], L. Glover[103], E. Goetz[9], R. Goetz[5], S. Gomes[96], G. González[2], J. M. Gonzalez Castro[22,23], A. Gopakumar[113], M. L. Gorodetsky[57], S. E. Gossan[1], M. Gosselin[28], R. Gouaty[7], A. Grado[4,114], C. Graef[43], M. Granata[25], A. Grant[43], S. Gras[14], C. Gray[44], G. Greco[65,66], A. C. Green[53], P. Groot[61], H. Grote[9], S. Grunewald[34], P. Gruning[26], G. M. Guidi[65,66], X. Guo[79], A. Gupta[82], M. K. Gupta[100], K. E. Gushwa[1], E. K. Gustafson[1], R. Gustafson[115], B. R. Hall[64], E. D. Hall[1], G. Hammond[43], M. Haney[113], M. M. Hanke[9], J. Hanks[44], C. Hanna[82], O. A. Hannuksela[88], J. Hanson[6], T. Hardwick[2], J. Harms[65,66], G. M. Harry[116], I. W. Harry[34], M. J. Hart[43], C.-J. Haster[85], K. Haughian[43], J. Healy[117], A. Heidmann[68], M. C. Heintze[6], H. Heitmann[62], P. Hello[26], G. Hemming[28], M. Hendry[43], I. S. Heng[43], J. Hennig[43], J. Henry[117], A. W. Heptonstall[1], M. Heurs[9,21], S. Hild[43], D. Hoak[28], D. Hofman[25], K. Holt[6], D. E. Holz[86], P. Hopkins[96], C. Horst[20], J. Hough[43], E. A. Houston[43], E. J. Howell[60], Y. M. Hu[9], E. A. Huerta[11], D. Huet[26], B. Hughey[33], S. Husa[97], S. H. Huttner[43], T. Huynh-Dinh[6], N. Indik[9], D. R. Ingram[44], R. Inta[80], G. Intini[32,91], H. N. Isa[43],

J.-M. Isac[68], M. Isi[1], B. R. Iyer[19], K. Izumi[44], T. Jacqmin[68], K. Jani[73], P. Jaranowski[118], S. Jawahar[119], F. Jiménez-Forteza[97], W. W. Johnson[2], D. I. Jones[120], R. Jones[43], R. J. G. Jonker[13], L. Ju[60], J. Junker[9], C. V. Kalaghatgi[96], V. Kalogera[95], S. Kandhasamy[6], G. Kang[36], J. B. Kanner[1], S. Karki[67], K. S. Karvinen[9], M. Kasprzack[2], M. Katolik[11], E. Katsavounidis[14], W. Katzman[6], S. Kaufer[21], K. Kawabe[44], F. Kéfélian[62], D. Keitel[43], A. J. Kemball[11], R. Kennedy[101], C. Kent[96], J. S. Key[121], F. Y. Khalili[57], I. Khan[16,17], S. Khan[9], Z. Khan[100], E. A. Khazanov[122], N. Kijbunchoo[44], Chunglee Kim[123], J. C. Kim[124], W. Kim[78], W. S. Kim[125], Y.-M. Kim[87,123], S. J. Kimbrell[73], E. J. King[78], P. J. King[44], R. Kirchhoff[9], J. S. Kissel[44], L. Kleybolte[31], S. Klimenko[5], P. Koch[9], S. M. Koehlenbeck[9], S. Koley[13], V. Kondrashov[1], A. Kontos[14], M. Korobko[31], W. Z. Korth[1], I. Kowalska[70], D. B. Kozak[1], C. Krämer[9], V. Kringel[9], B. Krishnan[9], A. Królak[126,127], G. Kuehn[9], P. Kumar[85], R. Kumar[100], S. Kumar[19], L. Kuo[83], A. Kutynia[126], S. Kwang[20], B. D. Lackey[34], K. H. Lai[88], M. Landry[44], R. N. Lang[20], J. Lange[117], B. Lantz[47], R. K. Lanza[14], A. Lartaux-Vollard[26], P. D. Lasky[109], M. Laxen[6], A. Lazzarini[1], C. Lazzaro[50], P. Leaci[32,91], S. Leavey[43], C. H. Lee[87], H. K. Lee[128], H. M. Lee[123], H. W. Lee[124], K. Lee[43], J. Lehmann[9], A. Lenon[37,38], M. Leonardi[90,104], N. Leroy[26], N. Letendre[7], Y. Levin[109], T. G. F. Li[88], A. Libson[14], T. B. Littenberg[129], J. Liu[60], R. K. L. Lo[88], N. A. Lockerbie[119], L. T. London[96], J. E. Lord[41], M. Lorenzini[16,17], V. Loriette[130], M. Lormand[6], G. Losurdo[23], J. D. Lough[9,21], C. O. Lousto[117], G. Lovelace[27], H. Lück[9,21], D. Lumaca[17,30], A. P. Lundgren[9], R. Lynch[14], Y. Ma[59], S. Macfoy[58], B. Machenschalk[9], M. MacInnis[14], D. M. Macleod[2], I. Magaña Hernandez[88], F. Magaña-Sandoval[41], L. Magaña Zertuche[41], R. M. Magee[82], E. Majorana[32], I. Maksimovic[130], N. Man[62], V. Mandic[42], V. Mangano[43], G. L. Mansell[24], M. Manske[20], M. Mantovani[28], F. Marchesoni[40,48], F. Marion[7], S. Márka[46], Z. Márka[46], C. Markakis[11], A. S. Markosyan[47], E. Maros[1], F. Martelli[65,66], L. Martellini[62], I. W. Martin[43], D. V. Martynov[14], K. Mason[14], A. Masserot[7], T. J. Massinger[1], M. Masso-Reid[43], S. Mastrogiovanni[32,91], A. Matas[42], F. Matichard[14], L. Matone[46], N. Mavalvala[14], R. Mayani[102], N. Mazumder[64], R. McCarthy[44], D. E. McClelland[24], S. McCormick[6], L. McCuller[14], S. C. McGuire[131], G. McIntyre[1], J. McIver[1], D. J. McManus[24], T. McRae[24], S. T. McWilliams[37,38], D. Meacher[82], G. D. Meadors[9,34], J. Meidam[13], E. Mejuto-Villa[8], A. Melatos[132], G. Mendell[44], R. A. Mercer[20], E. L. Merilh[44], M. Merzougui[62], S. Meshkov[1], C. Messenger[43], C. Messick[82], R. Metzdorff[68], P. M. Meyers[42], F. Mezzani[32,91], H. Miao[53], C. Michel[25], H. Middleton[53], E. E. Mikhailov[133], L. Milano[4,75], A. L. Miller[5], A. Miller[32,91], B. B. Miller[95], J. Miller[14], M. Millhouse[94], O. Minazzoli[62], Y. Minenkov[17], J. Ming[34], C. Mishra[134], S. Mitra[18], V. P. Mitrofanov[57], G. Mitselmakher[5], R. Mittleman[14], A. Moggi[23], M. Mohan[28], S. R. P. Mohapatra[14], M. Montani[65,66], B. C. Moore[105], C. J. Moore[12], D. Moraru[44], G. Moreno[44], S. R. Morriss[98], B. Mours[7], C. M. Mow-Lowry[53], G. Mueller[5], A. W. Muir[96], Arunava Mukherjee[9], D. Mukherjee[20], S. Mukherjee[98], N. Mukund[18], A. Mullavey[6], J. Munch[78], E. A. M. Muniz[41], P. G. Murray[43], K. Napier[73], I. Nardecchia[17,30], L. Naticchioni[32,91], R. K. Nayak[135], G. Nelemans[13,61], T. J. N. Nelson[6], M. Neri[54,55], M. Nery[9], A. Neunzert[115], J. M. Newport[116], G. Newton[43,152], K. K. Y. Ng[88], T. T. Nguyen[24], D. Nichols[61], A. B. Nielsen[9], S. Nissanke[13,61], A. Nitz[9], A. Noack[9], F. Nocera[28], D. Nolting[6], M. E. N. Normandin[98], L. K. Nuttall[41], J. Oberling[44], E. Ochsner[20], E. Oelker[14], G. H. Ogin[107], J. J. Oh[125], S. H. Oh[125], F. Ohme[9], M. Oliver[97], P. Oppermann[9], Richard J. Oram[6], B. O'Reilly[6], R. Ormiston[42], L. F. Ortega[5], R. O'Shaughnessy[117], D. J. Ottaway[78], H. Overmier[6], B. J. Owen[80], A. E. Pace[82], J. Page[129], M. A. Page[60], A. Pai[111], S. A. Pai[56], J. R. Palamos[67], O. Palashov[122], C. Palomba[32], A. Pal-Singh[31], H. Pan[83], B. Pang[59], P. T. H. Pang[88], C. Pankow[95], F. Pannarale[96], B. C. Pant[56], F. Paoletti[23], A. Paoli[28], M. A. Papa[9,20,34], H. R. Paris[47], W. Parker[6], D. Pascucci[43], A. Pasqualetti[28], R. Passaquieti[22,23], D. Passuello[23], B. Patricelli[23,136], B. L. Pearlstone[43], M. Pedraza[1], R. Pedurand[25,137], L. Pekowsky[41], A. Pele[6], S. Penn[138], C. J. Perez[44], A. Perreca[1,90,104], L. M. Perri[95], H. P. Pfeiffer[85], M. Phelps[43], O. J. Piccinni[32,91], M. Pichot[62], F. Piergiovanni[65,66], V. Pierro[8], G. Pillant[28], L. Pinard[25], I. M. Pinto[8], M. Pitkin[43], R. Poggiani[22,23], P. Popolizio[28], E. K. Porter[35], A. Post[9], J. Powell[43], J. Prasad[18], J. W. W. Pratt[33], V. Predoi[96], T. Prestegard[20], M. Prijatelj[9], M. Principe[8], S. Privitera[34], R. Prix[9], G. A. Prodi[90,104], L. G. Prokhorov[57], O. Puncken[9], M. Punturo[40], P. Puppo[32], M. Pürrer[34], H. Qi[20], J. Qin[60], S. Qiu[109], V. Quetschke[98], E. A. Quintero[1], R. Quitzow-James[67], F. J. Raab[44], D. S. Rabeling[24], H. Radkins[44], P. Raffai[51], S. Raja[56], C. Rajan[56], M. Rakhmanov[98], K. E. Ramirez[98], P. Rapagnani[32,91], V. Raymond[34], M. Razzano[22,23], J. Read[27], T. Regimbau[62], L. Rei[55], S. Reid[58], D. H. Reitze[1,5], H. Rew[133], S. D. Reyes[41], F. Ricci[32,91], P. M. Ricker[11], S. Rieger[9], K. Riles[115], M. Rizzo[117], N. A. Robertson[1,43], R. Robie[43], F. Robinet[26], A. Rocchi[17], L. Rolland[7], J. G. Rollins[1], V. J. Roma[67], R. Romano[3,4], C. L. Romel[44], J. H. Romie[6], D. Rosińska[52,139], M. P. Ross[140], S. Rowan[43], A Rüdiger[9], P. Ruggi[28], K. Ryan[44], M. Rynge[102], S. Sachdev[1], T. Sadecki[44], L. Sadeghian[20], M. Sakellariadou[141], L. Salconi[28], M. Saleem[111], F. Salemi[9], A. Samajdar[135], L. Sammut[109], L. M. Sampson[95], E. J. Sanchez[1], V. Sandberg[44], B. Sandeen[95], J. R. Sanders[41], B. Sassolas[25], B. S. Sathyaprakash[82,96], P. R. Saulson[41], O. Sauter[115], R. L. Savage[44], A. Sawadsky[21], P. Schale[67], J. Scheuer[95], E. Schmidt[33], J. Schmidt[9], P. Schmidt[1,61], R. Schnabel[31], R. M. S. Schofield[67], A. Schönbeck[31], A. Schönbeck[31], E. Schreiber[9], D. Schuette[9,21], B. W. Schulte[9], B. F. Schutz[9,96], S. G. Schwalbe[33], J. Scott[43], S. M. Scott[24], E. Seidel[11], D. Sellers[6], A. S. Sengupta[142], D. Sentenac[28], V. Sequino[17,30], A. Sergeev[122], D. A. Shaddock[24], T. J. Shaffer[44], A. A. Shah[129], M. S. Shahriar[95], L. Shao[34], B. Shapiro[47], P. Shawhan[72], A. Sheperd[20], D. H. Shoemaker[14], D. M. Shoemaker[73], K. Siellez[73], X. Siemens[20], M. Sieniawska[52], D. Sigg[44], A. D. Silva[15], A. Singer[1], L. P. Singer[76], A. Singh[9,21,34], R. Singh[2], A. Singhal[16,32], A. M. Sintes[97], B. J. J. Slagmolen[24], B. Smith[6], J. R. Smith[27], R. J. E. Smith[1], E. J. Son[125], J. A. Sonnenberg[20], B. Sorazu[43], F. Sorrentino[55], T. Souradeep[18], A. P. Spencer[43], A. K. Srivastava[100], A. Staley[46], M. Steinke[9], J. Steinlechner[31,43], S. Steinlechner[31], D. Steinmeyer[9,21], B. C. Stephens[20], R. Stone[98], K. A. Strain[43], G. Stratta[65,66], S. E. Strigin[57], R. Sturani[143], A. L. Stuver[6], T. Z. Summerscales[144], L. Sun[132], S. Sunil[100], P. J. Sutton[96], B. L. Swinkels[28], M. J. Szczepańczyk[33], M. Tacca[35], D. Talukder[67], D. B. Tanner[5], M. Tápai[112], A. Taracchini[34], J. A. Taylor[129], R. Taylor[1], T. Theeg[9], E. G. Thomas[53], M. Thomas[6], P. Thomas[44], K. A. Thorne[6], K. S. Thorne[59], E. Thrane[109],

S. Tiwari[16,90], V. Tiwari[96], K. V. Tokmakov[119], K. Toland[43], M. Tonelli[22,23], Z. Tornasi[43], C. I. Torrie[1], D. Töyrä[53], F. Travasso[28,40], G. Traylor[6], D. Trifirò[10], J. Trinastic[5], M. C. Tringali[90,104], L. Trozzo[23,145], K. W. Tsang[13], M. Tse[14], R. Tso[1], D. Tuyenbayev[98], K. Ueno[20], D. Ugolini[146], C. S. Unnikrishnan[113], A. L. Urban[1], S. A. Usman[96], K. Vahi[102], H. Vahlbruch[21], G. Vajente[1], G. Valdes[98], M. Vallisneri[59], N. van Bakel[13], M. van Beuzekom[13], J. F. J. van den Brand[13,71], C. Van Den Broeck[13], D. C. Vander-Hyde[41], L. van der Schaaf[13], J. V. van Heijningen[13], A. A. van Veggel[43], M. Vardaro[49,50], V. Varma[59], S. Vass[1], M. Vasúth[45], A. Vecchio[53], G. Vedovato[50], J. Veitch[53], P. J. Veitch[78], K. Venkateswara[140], G. Venugopalan[1], D. Verkindt[7], F. Vetrano[65,66], A. Viceré[65,66], A. D. Viets[20], S. Vinciguerra[53], D. J. Vine[58], J.-Y. Vinet[62], S. Vitale[14], T. Vo[41], H. Vocca[39,40], C. Vorvick[44], D. V. Voss[5], W. D. Vousden[53], S. P. Vyatchanin[57], A. R. Wade[1], L. E. Wade[81], M. Wade[81], R. Walet[13], M. Walker[2], L. Wallace[1], S. Walsh[20], G. Wang[16,66], H. Wang[53], J. Z. Wang[82], M. Wang[53], Y.-F. Wang[88], Y. Wang[60], R. L. Ward[24], J. Warner[44], M. Was[7], J. Watchi[92], B. Weaver[44], L.-W. Wei[9,21], M. Weinert[9], A. J. Weinstein[1], R. Weiss[14], L. Wen[60], E. K. Wessel[11], P. Wessels[9], T. Westphal[9], K. Wette[9], J. T. Whelan[117], B. F. Whiting[5], C. Whittle[109], D. Williams[43], R. D. Williams[1], A. R. Williamson[117], J. L. Willis[147], B. Willke[9,21], M. H. Wimmer[9,21], W. Winkler[9], C. C. Wipf[1], H. Wittel[9,21], G. Woan[43], J. Woehler[9], J. Wofford[117], K. W. K. Wong[88], J. Worden[44], J. L. Wright[43], D. S. Wu[9], G. Wu[6], W. Yam[14], H. Yamamoto[1], C. C. Yancey[72], M. J. Yap[24], Hang Yu[14], Haocun Yu[14], M. Yvert[7], M. Zanolin[33], T. Zelenova[28], J.-P. Zendri[50], M. Zevin[95], L. Zhang[1], M. Zhang[133], T. Zhang[43], Y.-H. Zhang[117], C. Zhao[60], M. Zhou[95], Z. Zhou[95], X. J. Zhu[60], M. E. Zucker[1,14], J. Zweizig[1]

LIGO Scientific Collaboration and Virgo Collaboration,

D. Steeghs[109,148], and L. Wang[148]

[1] LIGO, California Institute of Technology, Pasadena, CA 91125, USA
[2] Louisiana State University, Baton Rouge, LA 70803, USA
[3] Università di Salerno, Fisciano, I-84084 Salerno, Italy
[4] INFN, Sezione di Napoli, Complesso Universitario di Monte S.Angelo, I-80126 Napoli, Italy
[5] University of Florida, Gainesville, FL 32611, USA
[6] LIGO Livingston Observatory, Livingston, LA 70754, USA
[7] Laboratoire d'Annecy-le-Vieux de Physique des Particules (LAPP), Université Savoie Mont Blanc, CNRS/IN2P3, F-74941 Annecy, France
[8] University of Sannio at Benevento, I-82100 Benevento, Italy and INFN, Sezione di Napoli, I-80100 Napoli, Italy
[9] Albert-Einstein-Institut, Max-Planck-Institut für Gravitationsphysik, D-30167 Hannover, Germany
[10] The University of Mississippi, University, MS 38677, USA
[11] NCSA, University of Illinois at Urbana-Champaign, Urbana, IL 61801, USA
[12] University of Cambridge, Cambridge CB2 1TN, UK
[13] Nikhef, Science Park, 1098 XG Amsterdam, The Netherlands
[14] LIGO, Massachusetts Institute of Technology, Cambridge, MA 02139, USA
[15] Instituto Nacional de Pesquisas Espaciais, 12227-010 São José dos Campos, São Paulo, Brazil
[16] Gran Sasso Science Institute (GSSI), I-67100 L'Aquila, Italy
[17] INFN, Sezione di Roma Tor Vergata, I-00133 Roma, Italy
[18] Inter-University Centre for Astronomy and Astrophysics, Pune 411007, India
[19] International Centre for Theoretical Sciences, Tata Institute of Fundamental Research, Bengaluru 560089, India
[20] University of Wisconsin-Milwaukee, Milwaukee, WI 53201, USA
[21] Leibniz Universität Hannover, D-30167 Hannover, Germany
[22] Università di Pisa, I-56127 Pisa, Italy
[23] INFN, Sezione di Pisa, I-56127 Pisa, Italy
[24] OzGrav, Australian National University, Canberra, Australian Capital Territory 0200, Australia
[25] Laboratoire des Matériaux Avancés (LMA), CNRS/IN2P3, F-69622 Villeurbanne, France
[26] LAL, Univ. Paris-Sud, CNRS/IN2P3, Université Paris-Saclay, F-91898 Orsay, France
[27] California State University Fullerton, Fullerton, CA 92831, USA
[28] European Gravitational Observatory (EGO), I-56021 Cascina, Pisa, Italy
[29] Chennai Mathematical Institute, Chennai 603103, India
[30] Università di Roma Tor Vergata, I-00133 Roma, Italy
[31] Universität Hamburg, D-22761 Hamburg, Germany
[32] INFN, Sezione di Roma, I-00185 Roma, Italy
[33] Embry-Riddle Aeronautical University, Prescott, AZ 86301, USA
[34] Albert-Einstein-Institut, Max-Planck-Institut für Gravitationsphysik, D-14476 Potsdam-Golm, Germany
[35] APC, AstroParticule et Cosmologie, Université Paris Diderot, CNRS/IN2P3, CEA/Irfu, Observatoire de Paris, Sorbonne Paris Cité, F-75205 Paris Cedex 13, France
[36] Korea Institute of Science and Technology Information, Daejeon 34141, Korea
[37] West Virginia University, Morgantown, WV 26506, USA
[38] Center for Gravitational Waves and Cosmology, West Virginia University, Morgantown, WV 26505, USA
[39] Università di Perugia, I-06123 Perugia, Italy
[40] INFN, Sezione di Perugia, I-06123 Perugia, Italy
[41] Syracuse University, Syracuse, NY 13244, USA
[42] University of Minnesota, Minneapolis, MN 55455, USA
[43] SUPA, University of Glasgow, Glasgow G12 8QQ, UK
[44] LIGO Hanford Observatory, Richland, WA 99352, USA
[45] Wigner RCP, RMKI, H-1121 Budapest, Konkoly Thege Miklós út 29-33, Hungary
[46] Columbia University, New York, NY 10027, USA
[47] Stanford University, Stanford, CA 94305, USA
[48] Università di Camerino, Dipartimento di Fisica, I-62032 Camerino, Italy
[49] Università di Padova, Dipartimento di Fisica e Astronomia, I-35131 Padova, Italy
[50] INFN, Sezione di Padova, I-35131 Padova, Italy
[51] MTA Eötvös University, "Lendulet" Astrophysics Research Group, Budapest 1117, Hungary
[52] Nicolaus Copernicus Astronomical Center, Polish Academy of Sciences, 00-716, Warsaw, Poland
[53] University of Birmingham, Birmingham B15 2TT, UK

[54] Università degli Studi di Genova, I-16146 Genova, Italy
[55] INFN, Sezione di Genova, I-16146 Genova, Italy
[56] RRCAT, Indore MP 452013, India
[57] Faculty of Physics, Lomonosov Moscow State University, Moscow 119991, Russia
[58] SUPA, University of the West of Scotland, Paisley PA1 2BE, UK
[59] Caltech CaRT, Pasadena, CA 91125, USA
[60] OzGrav, University of Western Australia, Crawley, Western Australia 6009, Australia
[61] Department of Astrophysics/IMAPP, Radboud University Nijmegen, P.O. Box 9010, 6500 GL Nijmegen, The Netherlands
[62] Artemis, Université Côte d'Azur, Observatoire Côte d'Azur, CNRS, CS 34229, F-06304 Nice Cedex 4, France
[63] Institut de Physique de Rennes, CNRS, Université de Rennes 1, F-35042 Rennes, France
[64] Washington State University, Pullman, WA 99164, USA
[65] Università degli Studi di Urbino "Carlo Bo," I-61029 Urbino, Italy
[66] INFN, Sezione di Firenze, I-50019 Sesto Fiorentino, Firenze, Italy
[67] University of Oregon, Eugene, OR 97403, USA
[68] Laboratoire Kastler Brossel, UPMC-Sorbonne Universités, CNRS, ENS-PSL Research University, Collège de France, F-75005 Paris, France
[69] Carleton College, Northfield, MN 55057, USA
[70] Astronomical Observatory Warsaw University, 00-478 Warsaw, Poland
[71] VU University Amsterdam, 1081 HV Amsterdam, The Netherlands
[72] University of Maryland, College Park, MD 20742, USA
[73] Center for Relativistic Astrophysics and School of Physics, Georgia Institute of Technology, Atlanta, GA 30332, USA
[74] Université Claude Bernard Lyon 1, F-69622 Villeurbanne, France
[75] Università di Napoli 'Federico II", Complesso Universitario di Monte S.Angelo, I-80126 Napoli, Italy
[76] NASA Goddard Space Flight Center, Greenbelt, MD 20771, USA
[77] RESCEU, University of Tokyo, Tokyo, 113-0033, Japan
[78] OzGrav, University of Adelaide, Adelaide, South Australia 5005, Australia
[79] Tsinghua University, Beijing 100084, China
[80] Texas Tech University, Lubbock, TX 79409, USA
[81] Kenyon College, Gambier, OH 43022, USA
[82] The Pennsylvania State University, University Park, PA 16802, USA
[83] National Tsing Hua University, Hsinchu City, 30013 Taiwan, Republic of China
[84] Charles Sturt University, Wagga Wagga, New South Wales 2678, Australia
[85] Canadian Institute for Theoretical Astrophysics, University of Toronto, Toronto, Ontario M5S 3H8, Canada
[86] University of Chicago, Chicago, IL 60637, USA
[87] Pusan National University, Busan 46241, Korea
[88] The Chinese University of Hong Kong, Shatin, NT, Hong Kong
[89] INAF, Osservatorio Astronomico di Padova, Vicolo dell'Osservatorio 5, I-35122 Padova, Italy
[90] INFN, Trento Institute for Fundamental Physics and Applications, I-38123 Povo, Trento, Italy
[91] Università di Roma "La Sapienza", I-00185 Roma, Italy
[92] Université Libre de Bruxelles, Brussels 1050, Belgium
[93] Sonoma State University, Rohnert Park, CA 94928, USA
[94] Montana State University, Bozeman, MT 59717, USA
[95] Center for Interdisciplinary Exploration & Research in Astrophysics (CIERA), Northwestern University, Evanston, IL 60208, USA
[96] Cardiff University, Cardiff CF24 3AA, UK
[97] Universitat de les Illes Balears, IAC3—IEEC, E-07122 Palma de Mallorca, Spain
[98] The University of Texas Rio Grande Valley, Brownsville, TX 78520, USA
[99] Bellevue College, Bellevue, WA 98007, USA
[100] Institute for Plasma Research, Bhat, Gandhinagar 382428, India
[101] The University of Sheffield, Sheffield S10 2TN, UK
[102] University of Southern California Information Sciences Institute, Marina Del Rey, CA 90292, USA
[103] California State University, Los Angeles, 5151 State University Dr, Los Angeles, CA 90032, USA
[104] Università di Trento, Dipartimento di Fisica, I-38123 Povo, Trento, Italy
[105] Montclair State University, Montclair, NJ 07043, USA
[106] National Astronomical Observatory of Japan, 2-21-1 Osawa, Mitaka, Tokyo 181-8588, Japan
[107] Whitman College, 345 Boyer Avenue, Walla Walla, WA 99362 USA
[108] School of Mathematics, University of Edinburgh, Edinburgh EH9 3FD, UK
[109] OzGrav, School of Physics & Astronomy, Monash University, Clayton 3800, Victoria, Australia
[110] University and Institute of Advanced Research, Gandhinagar Gujarat 382007, India
[111] IISER-TVM, CET Campus, Trivandrum Kerala 695016, India
[112] University of Szeged, Dóm tér 9, Szeged 6720, Hungary
[113] Tata Institute of Fundamental Research, Mumbai 400005, India
[114] INAF, Osservatorio Astronomico di Capodimonte, I-80131, Napoli, Italy
[115] University of Michigan, Ann Arbor, MI 48109, USA
[116] American University, Washington, DC 20016, USA
[117] Rochester Institute of Technology, Rochester, NY 14623, USA
[118] University of Bia lystok, 15-424 Bia lystok, Poland
[119] SUPA, University of Strathclyde, Glasgow G1 1XQ, UK
[120] University of Southampton, Southampton SO17 1BJ, UK
[121] University of Washington Bothell, 18115 Campus Way NE, Bothell, WA 98011, USA
[122] Institute of Applied Physics, Nizhny Novgorod, 603950, Russia
[123] Seoul National University, Seoul 08826, Korea
[124] Inje University Gimhae, South Gyeongsang 50834, Korea
[125] National Institute for Mathematical Sciences, Daejeon 34047, Korea
[126] NCBJ, 05-400 Świerk-Otwock, Poland
[127] Institute of Mathematics, Polish Academy of Sciences, 00656 Warsaw, Poland
[128] Hanyang University, Seoul 04763, Korea
[129] NASA Marshall Space Flight Center, Huntsville, AL 35811, USA

[130] ESPCI, CNRS, F-75005 Paris, France
[131] Southern University and A&M College, Baton Rouge, LA 70813, USA
[132] OzGrav, University of Melbourne, Parkville, Victoria 3010, Australia
[133] College of William and Mary, Williamsburg, VA 23187, USA
[134] Indian Institute of Technology Madras, Chennai 600036, India
[135] IISER-Kolkata, Mohanpur, West Bengal 741252, India
[136] Scuola Normale Superiore, Piazza dei Cavalieri 7, I-56126 Pisa, Italy
[137] Université de Lyon, F-69361 Lyon, France
[138] Hobart and William Smith Colleges, Geneva, NY 14456, USA
[139] Janusz Gil Institute of Astronomy, University of Zielona Góra, 65-265 Zielona Góra, Poland
[140] University of Washington, Seattle, WA 98195, USA
[141] King's College London, University of London, London WC2R 2LS, UK
[142] Indian Institute of Technology, Gandhinagar Ahmedabad Gujarat 382424, India
[143] International Institute of Physics, Universidade Federal do Rio Grande do Norte, Natal RN 59078-970, Brazil
[144] Andrews University, Berrien Springs, MI 49104, USA
[145] Università di Siena, I-53100 Siena, Italy
[146] Trinity University, San Antonio, TX 78212, USA
[147] Abilene Christian University, Abilene, TX 79699, USA
[148] Department of Physics, University of Warwick, Gibbet Hill Road, Coventry CV4 7AL, UK

## Abstract

We present the results of a semicoherent search for continuous gravitational waves from the low-mass X-ray binary Scorpius X-1, using data from the first Advanced LIGO observing run. The search method uses details of the modeled, parametrized continuous signal to combine coherently data separated by less than a specified coherence time, which can be adjusted to trade off sensitivity against computational cost. A search was conducted over the frequency range 25–2000 Hz, spanning the current observationally constrained range of binary orbital parameters. No significant detection candidates were found, and frequency-dependent upper limits were set using a combination of sensitivity estimates and simulated signal injections. The most stringent upper limit was set at 175 Hz, with comparable limits set across the most sensitive frequency range from 100 to 200 Hz. At this frequency, the 95% upper limit on the signal amplitude $h_0$ is $2.3 \times 10^{-25}$ marginalized over the unknown inclination angle of the neutron star's spin, and $8.0 \times 10^{-26}$ assuming the best orientation (which results in circularly polarized gravitational waves). These limits are a factor of 3–4 stronger than those set by other analyses of the same data, and a factor of $\sim$7 stronger than the best upper limits set using data from Initial LIGO science runs. In the vicinity of 100 Hz, the limits are a factor of between 1.2 and 3.5 above the predictions of the torque balance model, depending on the inclination angle; if the most likely inclination angle of 44° is assumed, they are within a factor of 1.7.

## 1. Introduction

Rotating neutron stars (NSs) are the primary expected source of continuous, periodic gravitational waves (GWs) for ground-based GW detectors. Targets include known pulsars (Aasi et al. 2014a), non-pulsating NSs in supernova remnants (Wette et al. 2008; Abadie et al. 2010; Aasi et al. 2015a), and unknown isolated (Aasi et al. 2016; Abbott et al. 2016a) or binary NSs (Aasi et al. 2014b). A particularly promising source is an accreting NS in a low-mass X-ray binary (LMXB); accretion torque spins up the NS into the frequency band of the detectors, and the accretion can generate an asymmetric mass or current quadrupole that acts as the source for the GWs (Watts et al. 2008). An approximate equilibrium between the accretion spin-up and GW spin-down, as well as other spin-down torques can produce a signal that is nearly periodic in the NS's rest frame, and then Doppler-shifted due to the orbital motion of the NS and the motion of the detector on the surface of the Earth. Such an equilibrium scenario would produce a relation between the observed accretion-induced X-ray flux of the LMXB and the expected strength of the GWs. Scorpius X-1 (Sco X-1), the most luminous LMXB, is therefore a promising potential source of GWs (Papaloizou & Pringle 1978; Wagoner 1984; Bildsten 1998). Sco X-1 is presumed to consist of an NS of mass $\approx 1.4\,M_\odot$ in a binary orbit with a companion star of mass $\approx 0.4\,M_\odot$ (Steeghs & Casares 2002). Some of the parameters inferred from observations of the system are summarized in Table 1.

Several methods were used to search for Sco X-1 in data from the Initial LIGO science runs of 2002–2011: Abbott et al. (2007a) performed a fully coherent search (Jaranowski et al. 1998) on six hours of data from the second science run. Starting with the fourth science run, results for Sco X-1 were reported (Abbott et al. 2007b; Abadie et al. 2011) as part of a search for stochastic signals from isolated sky positions (Ballmer 2006). In the fifth science run, a search (Aasi et al. 2015b) was done for Doppler-modulated sidebands associated with the binary orbit (Messenger & Woan 2007; Sammut et al. 2014). In the sixth science run, Sco X-1 was included in a search (Aasi et al. 2014b) principally designed for unknown binary systems (Goetz & Riles 2011), and this method

[149] Deceased, March 2016.
[150] Deceased, March 2017.
[151] Deceased, February 2017.
[152] Deceased, December 2016.

**Table 1**
Observed Parameters of the LMXB Sco X-1

| Parameter | Value |
|---|---|
| R.A.[a] | $16^h19^m55\overset{s}{.}0850$ |
| Decl.[a] | $-15°38'24\overset{''}{.}9$ |
| Distance (kpc) | $2.8 \pm 0.3$ |
| Orbital inclination $i$[b] | $44° \pm 6°$ |
| $K_1$ (km s$^{-1}$)[c] | [10, 90] or [40, 90] |
| $T_{asc}$ (GPS s)[d] | $897753994 \pm 100$ |
| $P_{orb}$ (s)[d] | $68023.70 \pm 0.04$ |

**Notes.** Uncertainties are $1\sigma$ unless otherwise stated. There are uncertainties (relevant to the present search) in the projected velocity amplitude $K_1$ of the NS, the orbital period $P_{orb}$, and the time $T_{asc}$ at which the neutron star crosses the ascending node (moving away from the observer), measured in the solar system barycenter. The orbital eccentricity of Sco X-1 is believed to be small (Steeghs & Casares 2002; Wang 2017) and is ignored in this search. The inclusion of eccentric orbits would add two search parameters that are determined by the eccentricity and the argument of periapse (Messenger 2011; Leaci & Prix 2015).
[a] The sky position (as quoted in Abbott et al. 2007a and derived from Bradshaw et al. 1999) is determined to the microarcsecond, and therefore can be treated as known in the present search.
[b] The inclination $i$ of the orbit to the line of sight, from the observation of radio jets in Fomalont et al. (2001), is not necessarily the same as the inclination angle $\iota$ of the neutron star's spin axis, which determines the degree of polarization of the GW in Equation (1).
[c] The value of the projected orbital velocity $K_1$ is difficult to determine experimentally, and previous works used a value from Abbott et al. (2007a), which was derived with some assumptions from Steeghs & Casares (2002) and equivalent to $40 \pm 5$ km s$^{-1}$. The broader range listed here comes from Doppler tomography measurements and Monte Carlo simulations in Wang (2017), which show $K_1$ to be weakly determined beyond the constraint that 40 km s$^{-1} \lesssim K_1 \lesssim 90$ km s$^{-1}$. Preliminary results from Wang (2017) included the weaker constraint 10 km s$^{-1} \lesssim K_1 \lesssim 90$ km s$^{-1}$, which was used to determine the parameter range in Table 2.
[d] The time of ascension $T_{asc}$, at which the neutron star crosses the ascending node (moving away from the observer), measured in the solar system barycenter, is derived from the time of inferior conjunction of the companion given in Galloway et al. (2014) by subtracting $P_{orb}/4$. It corresponds to a time of 2008 June 17 16:06:20 UTC and can be propagated to other epochs by adding an integer multiple of $P_{orb}$, which results in increased uncertainty in $T_{asc}$ and correlations between $P_{orb}$ and $T_{asc}$; see Figure 1.
**References.** Bradshaw et al. (1999), Fomalont et al. (2001), Galloway et al. (2014), Wang (2017).

was subsequently improved to search directly for Sco X-1 (Meadors et al. 2016) and applied to Initial LIGO data (Meadors et al. 2017). A mock data challenge (Messenger et al. 2015) was conducted to compare several of the methods to search for Sco X-1, and the most sensitive (detecting all 50 simulated signals in the challenge, and 49 out of the 50 "training" signals) was the cross-correlation (CrossCorr) method (Dhurandhar et al. 2008; Whelan et al. 2015) used in the present analysis.[153]

The Advanced LIGO detectors (Aasi et al. 2015c) carried out their first observing run (O1) from 2015 September 12 to 2016 January 19 (Abbott et al. 2016b). Searches for transient signals were carried out in near-real time and resulted in the observation of the binary black hole (BBH) mergers GW 150914 (Abbott et al. 2016c) and GW 151226 (Abbott et al. 2016d), and the possible BBH merger LVT 151012 (Abbott et al. 2016b), as well as upper limits on the rates and strengths of other sources (Abbott et al. 2016e, 2017a, 2017b). Searches for persistent stochastic or periodic sources were conducted using data from the full duration of the run and include searches for isotropic and anisotropic stochastic signals (Abbott et al. 2017c, 2017d) and a variety of known and unknown NSs (Abbott et al. 2017e). So far, two analyses including searches for GWs from Sco X-1 besides the current one have been released: Abbott et al. (2017d) included the direction of Sco X-1 in their directed unmodeled search for persistent GWs, and Abbott et al. (2017f) performed a directed search for Sco X-1 using a hidden Markov model.

[153] The CrossCorr analysis was carried out in "self-blinded" mode without knowledge of the simulated signal parameters, after the nominal end of the challenge.

## 2. Model of GWs from Sco X-1

The modeled GW signal from a rotating NS consists of a "plus" polarization component, $h_+(t) = A_+ \cos[\Phi(t)]$, and a "cross" polarization component, $h_\times(t) = A_\times \sin[\Phi(t)]$. The signal recorded in a particular detector will be a linear combination of $h_+$ and $h_\times$ determined by the detector's orientation as a function of time. The two polarization amplitudes are

$$A_+ = h_0 \frac{1 + \cos^2\iota}{2} \quad \text{and} \quad A_\times = h_0 \cos\iota, \tag{1}$$

where $h_0$ is an intrinsic amplitude related to the NS's ellipticity, moment of inertia, spin frequency, and distance; and $\iota$ is the inclination of the NS's spin to the line of sight. (For an NS in a binary, this may or may not be related to the inclination $i$ of the binary orbit.) If $\iota = 0°$ or $180°$, $A_\times = \pm A_+$, and gravitational radiation is circularly polarized. If $\iota = 90°$, $A_\times = 0$, it is linearly polarized. The general case, elliptical polarization, has $0 < |A_\times| < A_+$. Many search methods are sensitive to the combination

$$(h_0^{\mathrm{eff}})^2 = \frac{A_+^2 + A_\times^2}{2} = h_0^2 \frac{[(1 + \cos^2\iota)/2]^2 + [\cos\iota]^2}{2}, \tag{2}$$

which is equal to $h_0^2$ for circular polarization and $h_0^2/8$ for linear polarization (Messenger et al. 2015; note that this differs by a factor of 2.5 from the definition of $(h_0^{\mathrm{eff}})^2$ in Whelan et al. 2015).

It has been suggested (Papaloizou & Pringle 1978; Wagoner 1984; Bildsten 1998) that an LMXB may be in an equilibrium state where the spin-up due to accretion is due to the spin-down due to GWs. In that case, the GW amplitude can be related to the accretion rate, as inferred from the X-ray flux $F_X$ (Watts et al. 2008):

$$h_0 \approx 3 \times 10^{-27} \left(\frac{F_X}{10^{-8}\,\mathrm{erg\,cm^{-2}\,s^{-1}}}\right)^{1/2} \left(\frac{\nu_s}{300\,\mathrm{Hz}}\right)^{-1/2} \times \left(\frac{R}{10\,\mathrm{km}}\right)^{3/4} \left(\frac{M}{1.4\,M_\odot}\right)^{-1/4}.$$

For Sco X-1, using the observed X-ray flux $F_X = 3.9 \times 10^{-7}$ erg cm$^{-2}$ s$^{-1}$ from Watts et al. (2008), and assuming that the GW frequency $f_0$ is twice the spin frequency $\nu_s$ (as would be the case for GWs generated by triaxiality in the NS), the torque

balance value is

$$h_0 \approx 3.4 \times 10^{-26}\left(\frac{f_0}{600\ \mathrm{Hz}}\right)^{-1/2}. \tag{4}$$

Recent works (Haskell et al. 2015a, 2015b) have cast doubt on the ubiquity of the GW torque balance scenario in light of other spin-down mechanisms; the torque balance level remains an important benchmark for search sensitivity, and the detection or non-detection at or below that level would provide insight into the behavior of accreting NSs.

## 3. CrossCorr Search Method

The CrossCorr method was presented in Dhurandhar et al. (2008) and refined for application to Sco X-1 in Whelan et al. (2015). It was applied to simulated Advanced LIGO data in a mock data challenge (Messenger et al. 2015; Y. Zhang et al. 2017, in preparation). It was originally developed as a model-based improvement of the directional stochastic search of Ballmer (2006), which has been used to set limits on gravitational radiation from specific sky directions including Sco X-1 (Abbott et al. 2007b; Abadie et al. 2011). The method allows data to be correlated up to an adjustable coherence time $T_{\max}$. The data are split into segments of length $T_{\mathrm{sft}}$ between 240 and 1400 s (depending on frequency) and Fourier transformed. In a given data segment or short Fourier transform (SFT), the signal is expected to be found in a particular Fourier bin (or bins, considering the effects of spectral leakage). The signal bins are determined by the intrinsic frequency and the expected Doppler shift, which is in turn determined by the time and detector location, as well as the assumed orbital parameters of the LMXB. If the SFTs are labelled by the index $K$, $L$, etc., which encodes both the detector in question and the time of the SFT, and $z_K$ is the appropriately normalized Fourier data in the bin(s) of interest, the CrossCorr statistic has the form

$$\rho = \sum_{KL\in\mathcal{P}} (W_{KL} z_K^* z_L + W_{KL}^* z_K z_L^*). \tag{5}$$

This includes the product of the data from SFTs $K$ and $L$, where $KL$ is in a list of allowed pairs $\mathcal{P}$, defined by $K < L$ and $|T_K - T_L| \leqslant T_{\max}$, i.e., the times of the two different data segments should differ by no more than some specified lag time $T_{\max}$, which we also refer to as the coherence time. The complex weighting factors $W_{KL}$ are chosen (according to Equations (2.33)–(2.36) and (3.5) of Whelan et al. 2015) to maximize the expected statistic value subject to the normalization $\mathrm{Var}(\rho) = 1$. The expected statistic value is then

$$E[\rho] = (h_0^{\mathrm{eff}})^2 \vartheta, \tag{6}$$

where

$$\vartheta \approx 0.903 \sqrt{N_{\mathrm{det}}^2 T_{\mathrm{obs}} T_{\max} \left\langle \frac{4(\Gamma_{KL}^{\mathrm{ave}})^2}{S_K S_L} \right\rangle_{KL\in\mathcal{P}}}, \tag{7}$$

(this is the quantity called $\varrho^{\mathrm{ave}}$ in Whelan et al. 2015) and $h_0^{\mathrm{eff}}$ is the combination of $h_0$ and $\cos\iota$ defined in Equation (2), $S_K$ is constructed from the noise power spectrum and $\Gamma_{KL}^{\mathrm{ave}}$ from the antenna patterns for detectors $K$ and $L$ at the appropriate times, $N_{\mathrm{det}}$ is the number of detectors participating in the search, $T_{\mathrm{obs}}$ is the observing time per detector, and the factor of 0.903 arises from spectral leakage, assuming we consider contributions from all Fourier bins. (See Equation (3.19) of Whelan et al. 2015 for more details.) Increasing $T_{\max}$ increases the sensitivity of the search, but also increases the computing cost. In order to maximize the chance for a potential detection, a range of choices for $T_{\max}$ was used for different values of signal frequency and orbital parameters. The method used longer coherence times in regions of parameter space where (1) the detectable signal level given the frequency-dependent instrumental noise was closer to the expected signal strength from torque balance, (2) the cost of the search was lower due to template spacing, i.e., at lower frequencies and $a \sin i$ values, or (3) the signal had higher prior probability of being found, i.e., closer to the most likely value of $T_{\mathrm{asc}}$. This is illustrated in Figure 2. The full set of coherence times used ranges from 25,290 s for 25–50 Hz (for the most likely $T_{\mathrm{asc}}$ and smallest $a \sin i$ values) to 240 s at frequencies above 1200 Hz.

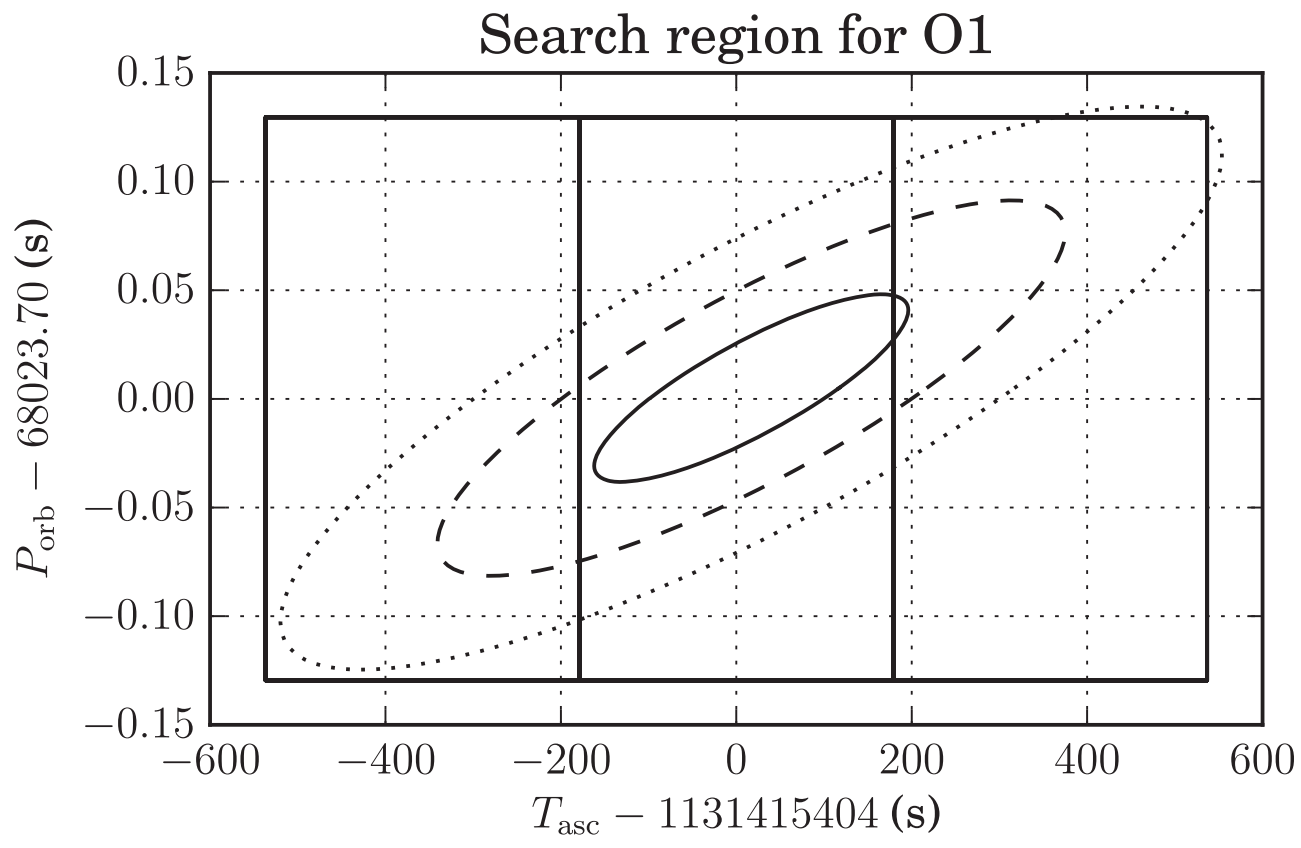

**Figure 1.** Range of search parameters $T_{\mathrm{asc}}$ and $P_{\mathrm{orb}}$. The ellipses show curves of constant prior probability corresponding to $1\sigma$, $2\sigma$, and $3\sigma$ (containing 39.3%, 86.5%, and 98.9% of the prior probability, respectively), including the effect of correlations arising from the propagation of the $T_{\mathrm{asc}}$ estimate from Table 1 to the mid-run value in Table 2. The search region is chosen to include the $3\sigma$ ellipse, with the range of $T_{\mathrm{asc}}$ within $\pm 1\sigma$ receiving a deeper search, as illustrated in Figure 2. The inner and outer regions contain 68.0% and 99.5% of the prior probability, respectively. Note that the apparent inefficiency in searching unlikely regions of $T_{\mathrm{asc}}$–$P_{\mathrm{orb}}$ space is mitigated by the fact that the search does not typically resolve $P_{\mathrm{orb}}$, resulting in only one value being included in the search.

The search was performed using a bank of template signals laid out in hypercubic lattice in the signal parameters of intrinsic frequency $f_0$, projected semimajor axis $a \sin i$, time of ascension $T_{\mathrm{asc}}$, and (where appropriate) orbital period $P_{\mathrm{orb}}$. The range of values in each direction, motivated by Table 1 and Figure 1, is shown in Table 2. The lattice spacing for the initial search was chosen to correspond to a nominal metric mismatch (fractional loss of signal-to-noise ratio (S/N) associated with a one-lattice-spacing offset in a given direction, assuming quadratic approximation) of 25% in each of the four parameters, using the metric computed in Whelan et al. (2015). The lattice was constructed (and spacing computed) for each of the 18 orbital parameter space cells shown in Figure 2 in each 0.05 Hz-wide frequency band. This resulted in a total of $\sim 9 \times 10^5$–$2 \times 10^8$ detection statistics per 0.05 Hz, as detailed in Table 3.

## 4. Follow-up of Candidates

Although the detection statistic $\rho$ is normalized to have zero mean and unit variance in Gaussian noise, the trials factor associated with the large number of templates at different points in parameter space results in numerous candidates with $\rho \gtrsim 6$. A follow-up was performed whenever $\rho$ exceeded a threshold of 6.5 for $25\ \mathrm{Hz} < f_0 < 400\ \mathrm{Hz}$, 6.2 for $400\ \mathrm{Hz} < f_0 < 600\ \mathrm{Hz}$, and 6.0 for $600\ \mathrm{Hz} < f_0 < 2000\ \mathrm{Hz}$. These thresholds were chosen in light of the number of templates searched (cf. Table 3) as a function of frequency. For each 5 Hz band, the threshold at which the expected number of Gaussian outliers was 0.1 (Figure 3). For simplicity, the three thresholds (6.5, 6.2, and 6.0) were chosen to be close to or slightly below these threshold values. As a result, the number of expected Gaussian outliers per 5 Hz was between 0.06 and 0.92. Table 3 shows the total expected number of outliers in each range of frequencies under the Gaussian assumption. Since the noise was not Gaussian, the actual number of signals followed up was significantly larger, as also shown in Table 3.

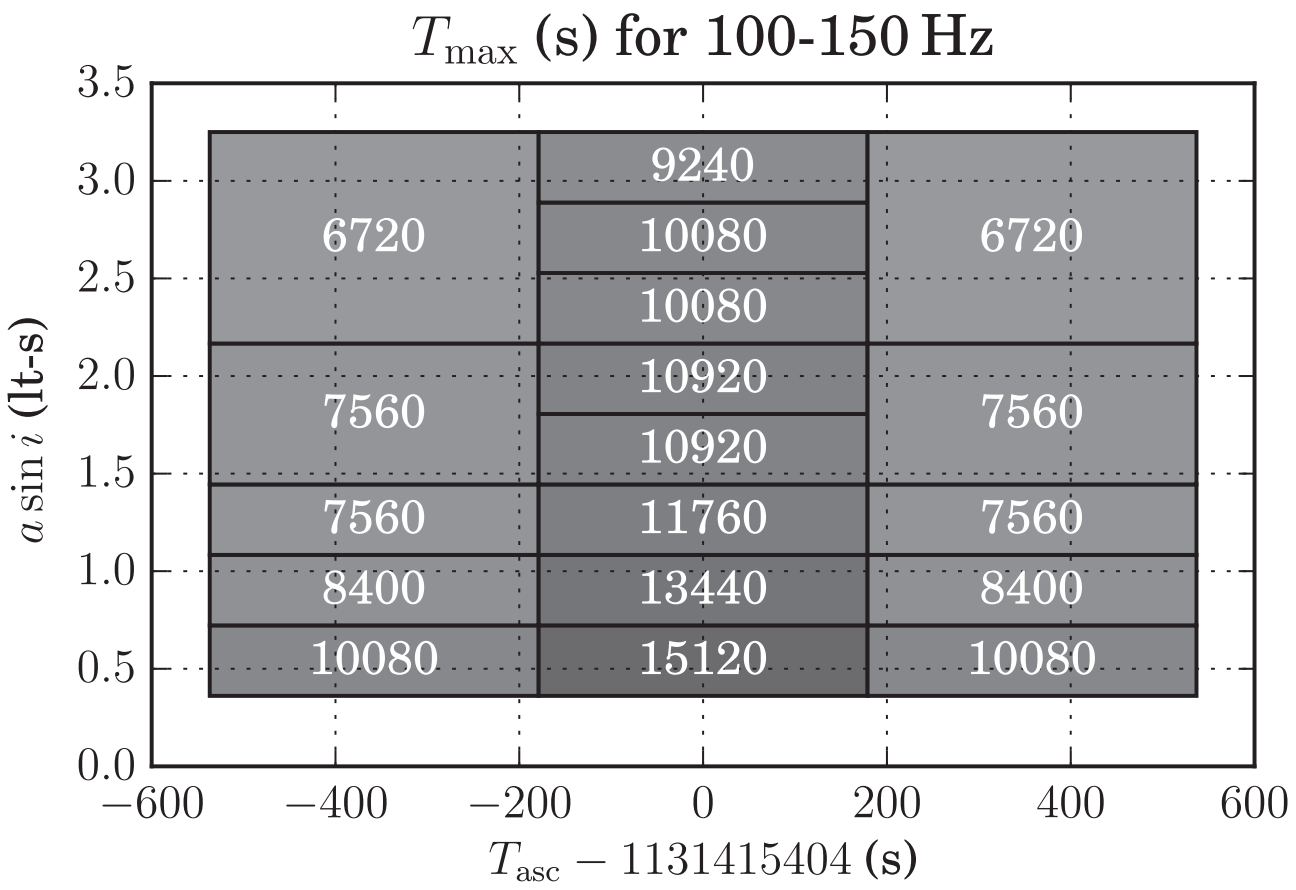

**Figure 2.** Example of coherence times $T_{max}$, in seconds, chosen as a function of the orbital parameters of the NS. Increasing coherence time improves the sensitivity but increases the computational cost of the search. The values are chosen to roughly optimize the search (by maximizing the detection probability at fixed computing cost subject to some arbitrary assumptions about the prior on $h_0$) assuming a uniform prior on the projected semimajor axis $a \sin i$ and a Gaussian prior on the time of ascension $T_{asc}$. Longer coherence times are used for more likely values of $T_{asc}$ (within $\pm 1\sigma$ of the mode of the prior distribution) and for smaller values of $a \sin i$ (where the parameter space metric of Whelan et al. 2015 implies a coarser resolution in $T_{asc}$ and reduced computing cost).

The follow-up procedure was as follows:

1. Data contaminated by known monochromatic noise features ("lines") in each detector were excluded from the search from the start. In most cases, the time-dependent orbital Doppler modulation of the expected signal meant that a narrow line only affected data relevant to a subset of the SFTs from the run. Pairs involving these SFTs needed to be excluded from the sum in Equation (5) and the normalization in Equation (7). The impact of this is illustrated in Figure 6 (in Appendix A), which shows the reduction in the sensitivity $\vartheta$ from the omission of pairs from Equation (7).
2. Because a strong signal generally led to elevated statistic values over a range of frequencies, all of the candidates within 0.02 Hz of a local maximum were "clustered" together, with the location of the maximum determining the parameters of the candidate signal. These are known as the "level 0" results.
3. A "refinement" search was performed in a $13 \times 13 \times 13 \times 13$ grid in $f_0$, with the same $T_{max}$ as the original search, and $a \sin i$, $T_{asc}$, and $P_{orb}$ centered on the original candidate, with a grid spacing chosen to be one-third of the original spacing (with appropriate modifications for $P_{orb}$ depending on whether that parameter was resolved in the original search). This procedure produces a grid that covers $\pm 2$ grid spacings of the original grid and has a mismatch of approximately $25\% \times (1/3)^2 \approx 2.8\%$. The results of this refinement stage are known as "level 1."
4. A deeper follow-up was done on the level 1 results, with $T_{max}$ increased to $4\times$ its original value. According to the the theoretical expectation in Equation (7), this should approximately double the statistic value $\rho$ for a true signal. Since this increase in coherence time also produces a finer parameter space resolution, the density of the grid was again increased by a further factor of 3 in each direction (resulting in a mismatch of approximately $25\% \times (1/3)^2 \times (4/3)^2 \approx 4.9\%$),[154] and the size of the grid was $13 \times 13 \times 13 \times 13$. The results of this follow-up stage are known as "level 2." Signals whose detection statistic $\rho$ decreases at this stage are dropped from the follow-up.
5. Surviving level 2 results were followed up by once again quadrupling the coherence time $T_{max}$ to $16\times$ the original value, and increasing the density by a factor of 3 in each direction, for an approximate mismatch of $25\% \times (1/3)^2 \times (4/3)^2 \times (4/3)^2 \approx 8.8\%$. Again, true signals are expected to approximately double their statistic values, and the grid is modified as at level 2. The results of this round of follow-up are known as "level 3."
6. Unknown instrumental lines in a single detector are likely to produce strong correlations between SFTs from that detector. To check for this, at each stage of follow-up, level 1 and beyond, the CrossCorr statistic $\rho_{HH}$ was calculated using only data from LIGO Hanford Observatory (LHO), and the statistic $\rho_{LL}$ using only data from LIGO Livingston Observatory (LLO). If we write $\rho_{HL}$ as

**Table 2**
Parameters Used for the Cross-correlation Search

| Parameter | Range |
|---|---|
| $f_0$ (Hz) | [25, 2000] |
| $a \sin i$ (lt-s)[a] | [0.36, 3.25] |
| $T_{asc}$ (GPS s)[b] | $1131415404 \pm 3 \times 179$ |
| $P_{orb}$ (s) | $68023.70 \pm 3 \times 0.04$ |

**Notes.** Ranges for $T_{asc}$ and $P_{orb}$ are chosen to cover $\pm 3\sigma$ of the observational uncertainties, as illustrated in Figure 1.
[a] The range for the projected semimajor axis, $a \sin i = K_1 P_{orb}/(2\pi)$, in light-seconds was taken from the constraint $K_1 \in [10, 90]\ \mathrm{km\ s^{-1}}$, which was the preliminary finding of Wang (2017) available at the time the search was constructed. Note that this range of $a \sin i$ values is broader than that used in previous analyses, which assumed a value from Abbott et al. (2007a) of 1.44 lt-s with a $1\sigma$ uncertainty of 0.18 lt-s.
[b] This value for the time of ascension has been propagated forward by 3435 orbits from the value in Table 1, and corresponds to a time of 2015 November 13 02:03:07 UTC, near the middle of the O1 run. (This is useful when constructing the lattice to search over the orbital parameter space, as noted in Whelan et al. 2015.) The increase in uncertainty is due to the uncertainty in $P_{orb}$.

[154] Note that the increased mismatch means that the highest S/N may not quite double, even for a true signal. As Figure 4 shows, simulated signals still show significant increases in S/N at levels 2 and 3 of the follow-up.

**Table 3**
Summary of Numbers of Templates and Candidates

| Min $f_0$ (Hz) | Max $f_0$ (Hz) | Min $T_{\max}$ (s) | Max $T_{\max}$ (s) | $\rho$ Threshold | Number of Templates | Expected Gauss False Alarms[a] | Level 0[b] | Level 1[c] | Level 2[d] | Level 3[e] |
|---|---|---|---|---|---|---|---|---|---|---|
| 25 | 50 | 10,080 | 25,920 | 6.5 | $1.58 \times 10^{10}$ | 0.6 | 269 | 212 | 62 | 6 |
| 50 | 100 | 8160 | 19,380 | 6.5 | $7.96 \times 10^{10}$ | 3.2 | 499 | 473 | 209 | 14 |
| 100 | 150 | 6720 | 15,120 | 6.5 | $1.51 \times 10^{11}$ | 6.1 | 605 | 571 | 304 | 29 |
| 150 | 200 | 5040 | 11,520 | 6.5 | $1.62 \times 10^{11}$ | 6.5 | 456 | 432 | 260 | 35 |
| 200 | 300 | 2400 | 6600 | 6.5 | $1.33 \times 10^{11}$ | 5.3 | 220 | 194 | 87 | 29 |
| 300 | 400 | 1530 | 4080 | 6.5 | $6.62 \times 10^{10}$ | 2.7 | 254 | 216 | 23 | 10 |
| 400 | 600 | 360 | 1800 | 6.5 | $1.62 \times 10^{10}$ | 0.6 | 88 | 26 | 2 | 1 |
| 600 | 800 | 360 | 720 | 6.2 | $5.80 \times 10^{9}$ | 1.6 | 78 | 15 | 2 | 2 |
| 800 | 1200 | 300 | 300 | 6.0 | $1.18 \times 10^{10}$ | 11.7 | 145 | 134 | 3 | 0 |
| 1200 | 2000 | 240 | 240 | 6.0 | $3.12 \times 10^{10}$ | 30.8 | 442 | 107 | 6 | 1 |

**Notes.** For each range of frequencies, this table shows the minimum and maximum coherence time $T_{\max}$ used for the search, across the different orbital parameter space cells (see Figure 2), the threshold in signal-to-noise ratio (S/N) $\rho$ used for follow-up, the total number of templates, and the number of candidates at various stages of the process. (See Section 4 for detailed description of the follow-up procedure.)
[a] This is the number of candidates that would be expected in Gaussian noise, given the number of templates and the follow-up threshold.
[b] This is actual number of candidates (after clustering) that crossed the S/N threshold and were followed up.
[c] This is the number of candidates remaining after refinement. All of the candidates "missing" at this stage have been removed by the single-detector veto for unknown lines.
[d] This is the number of candidates remaining after each one has been followed up with a $T_{\max}$ equal to 4× the original $T_{\max}$ for that candidate. (True signals should approximately double their S/N; any candidates whose S/N goes down have been dropped.) All of the signals present at this stage are shown in Figure 4, which also shows the behavior of the search on simulated signals injected in software.
[e] This is the number of candidates remaining after $T_{\max}$ has been increased to 16× its original value.

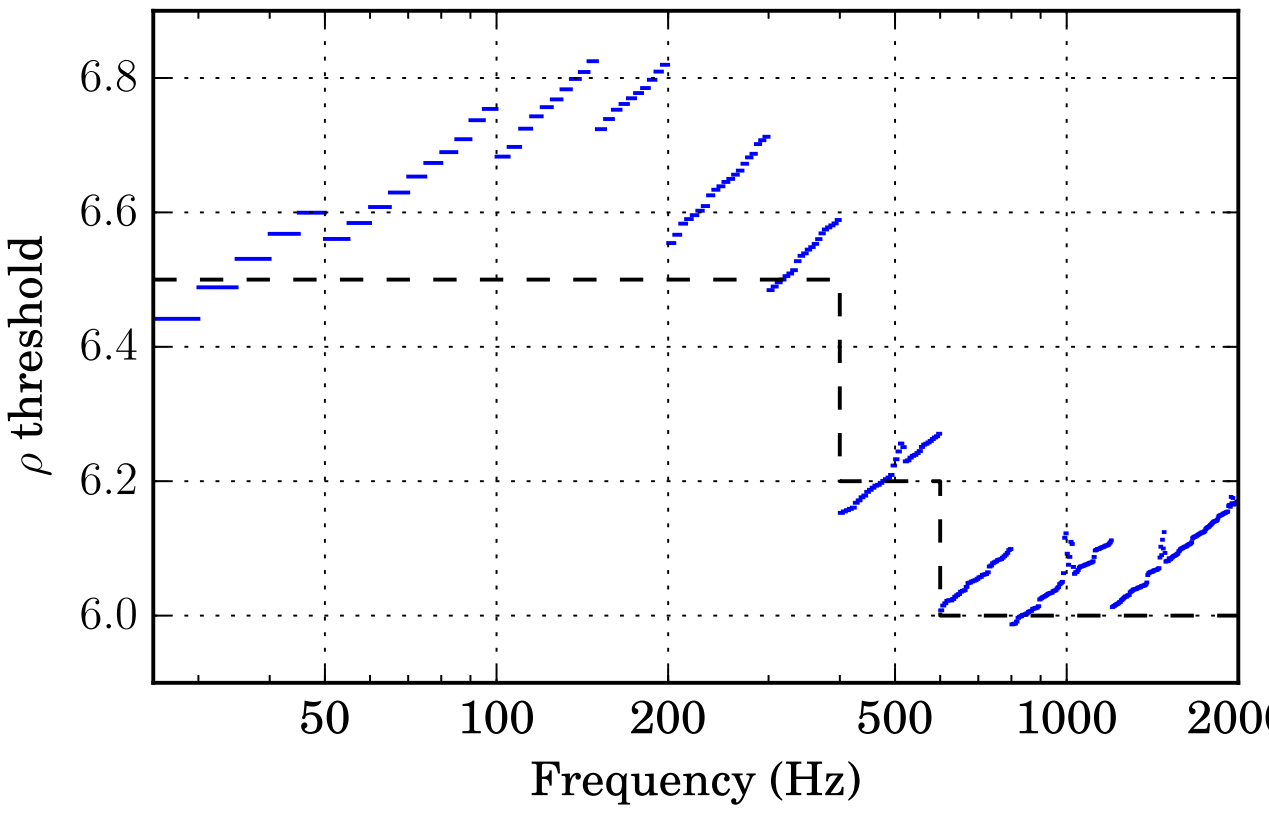

**Figure 3.** Selection of follow-up threshold as a function of frequency. If the data contained no signal and only Gaussian noise, each template in parameter space would have some chance of producing a statistic value exceeding a given threshold. Within each 0.05 Hz frequency band, the total number of templates was computed and used to find the threshold at which the expected number of Gaussian outliers above that value would be 0.1 (short blue lines). For simplicity, the actual follow-up threshold was chosen near or below that level, producing thresholds of 6.5 for 25 Hz $< f_0 <$ 400 Hz, 6.2 for 400 Hz $< f_0 <$ 600 Hz, and 6.0 for 600 Hz $< f_0 <$ 2000 Hz (black dashed line). Note that the large number of non-Gaussian outliers (cf. Table 3) makes the Gaussian follow-up level an imprecise tool in any event.

the statistic constructed using only pairs of one SFT from LHO and one from LLO, the overall statistic can be written (cf. Equations (2.36), (3.6) and (3.7) of Whelan et al. 2015) as

$$\rho = \frac{\vartheta_{\mathrm{HH}}\rho_{\mathrm{HH}} + \vartheta_{\mathrm{LL}}\rho_{\mathrm{LL}} + \vartheta_{\mathrm{HL}}\rho_{\mathrm{HL}}}{\vartheta}, \tag{8}$$

where

$$\vartheta = \sqrt{\vartheta_{\mathrm{HH}}^2 + \vartheta_{\mathrm{LL}}^2 + \vartheta_{\mathrm{HL}}^2}. \tag{9}$$

Since, for example, $E[\rho] = (h_0^{\mathrm{eff}})^2\vartheta > (h_0^{\mathrm{eff}})^2\vartheta_{\mathrm{HH}} = E[\rho_{\mathrm{HH}}]$, we expect true signals to have higher overall detection statistics $\rho$ than the single-detector statistics $\rho_{\mathrm{HH}}$ and $\rho_{\mathrm{LL}}$. We therefore veto any candidate for which $\rho_{\mathrm{HH}} > \rho$ or $\rho_{\mathrm{LL}} > \rho$ at any level of follow-up. This is responsible for the reduction of candidates from level 0 to level 1 seen in Table 3.

A total of 127 candidates survive level 3 of the follow-up. To check whether any of them represent convincing detection candidates, we plot in Figure 4 the ratio by which the S/N increases from level 1 to level 2, and from level 2 to level 3. We also plot the corresponding ratios for all of the candidates surviving level 2 (the 16× original $T_{\max}$ follow-up is not available for candidates that fail level 2), and also for the simulated signal injections described in Appendix A. We see that none of the candidates come close to doubling their S/N at either stage; in fact, none of them even double their S/N from level 1 to level 3. We empirically assess the follow-up procedure with the injections, and find that their S/Ns generally increase by slightly less than the naïvely expected factor of 2 (perhaps because of the increasing mismatch at later follow-up levels). We do see that the injected signals (at least those that survive level 2 follow-up and appear on the plot) nearly all increase their S/N noticeably more than any of the candidates from the search. Also note that of the 666 injected signals (out of 754) that produced $\rho$ values above their respective thresholds, 652 survived all levels of follow-up. (There were four vetoed at level 1, four at level 2, and six at level 3 of the follow-up.) All but a handful of those 652 (between one and four, depending on the stringency of the criterion) are well-separated from the bulk of the search results in Figure 4. We thus conclude that our follow-up procedure is relatively robust, and that there are no convincing detection candidates from the search.

The signal model in this search assumes that the GW frequency $f_0$ in the NS's reference frame is constant. In practice, the equilibrium in an LMXB will be only approximate, and the intrinsic frequency will vary stochastically with time. Whelan et al. (2015) estimated the effect of spin wandering under a simplistic random-walk model in which the GW frequency underwent a net spin-up or spin-down of magnitude $|\dot{f}|_{\text{drift}}$, changing on a timescale $T_{\text{drift}}$. The fractional loss of S/N was estimated as

$$\frac{E[\rho]^{\text{ideal}} - E[\rho]}{E[\rho]^{\text{ideal}}} \approx \frac{\pi^2}{6} T_{\text{run}} T_{\text{drift}} |\dot{f}|^2_{\text{drift}} T^2_{\max}, \tag{10}$$

where $T_{\text{run}}$ is the duration of the observing run from the start to end, not considering duty factors (in contrast to the $T_{\text{obs}}$ appearing in Equation (7)) or numbers of detectors. To give an illustration of the possible impacts of spin wandering on the present search, we make reference to the values of $|\dot{f}|_{\text{drift}} = 10^{-12}$ Hz s$^{-1}$ and $T_{\text{drift}} = 10^6$ s. These are conservative upper limits on how fast the signal can drift, based on Bildsten (1998). Similar values have been used in the first Sco X-1 mock data challenge (Messenger et al. 2015) and other work on Sco X-1 (Leaci & Prix 2015; Whelan et al. 2015).[155]

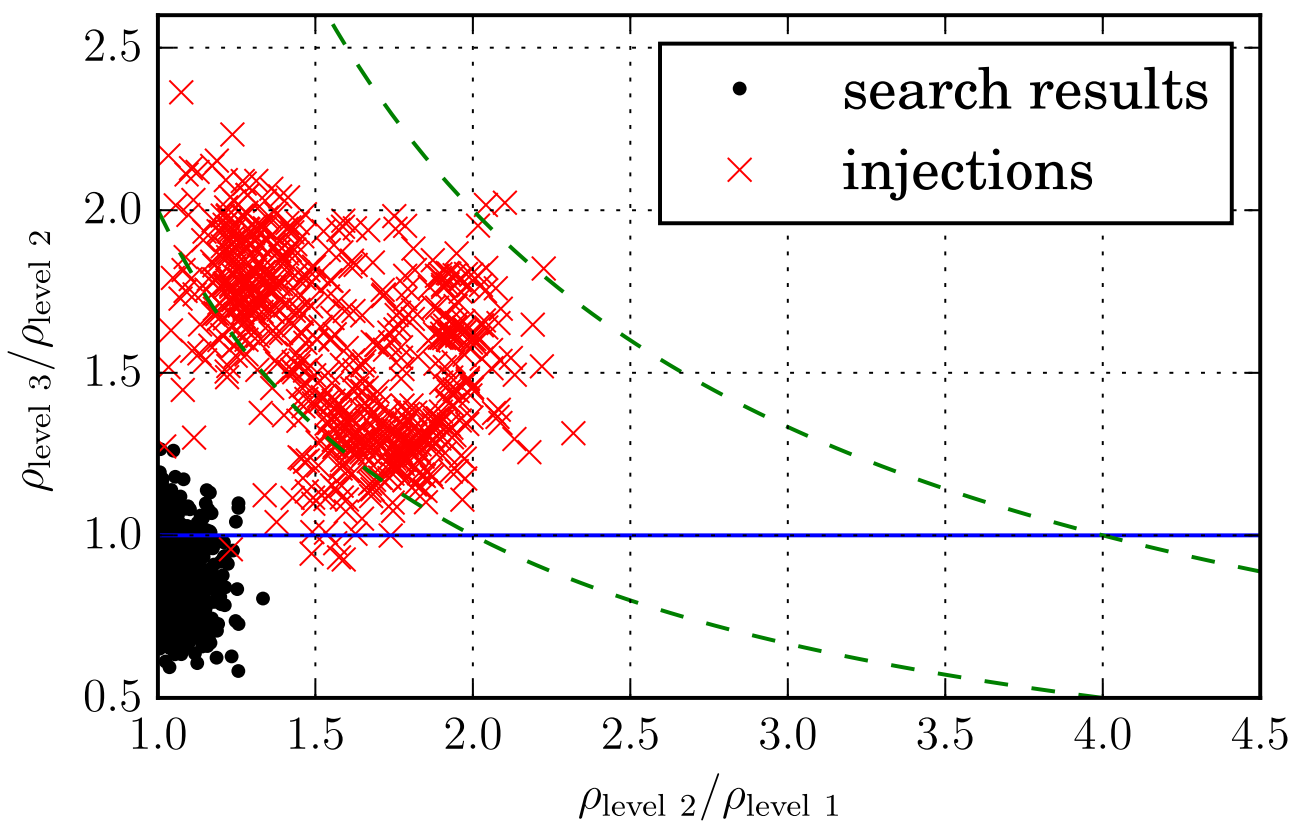

**Figure 4.** Ratios of follow-up statistics for search candidates and simulated signals. This plot shows all of the candidates that survived to level 2 of the follow-up (see Section 4 and Table 3), both from the main search and from the analysis of the simulated signal injections described in Appendix A. It shows the ratios of the S/N $\rho$ after follow-up level 1 (at the original coherence time $T_{\max}$), level 2 (at $4\times$ the original coherence time), and level 3 (at $16\times$ the original coherence time). The green dashed lines are at constant values of $\rho_{\text{level3}}/\rho_{\text{level 1}}$ equal to 2 and 4, respectively. There are no points with $\rho_{\text{level 2}}/\rho_{\text{level1}} < 1$, because those candidates do not survive level 2 follow-up and are therefore not subjected to level 3 follow-up. From Equation (5) and Equation (7), the naïve expectation is that the S/N will roughly double each time $T_{\max}$ is quadrupled. Empirically, the follow-ups of injections do not show exactly that relationship, but the vast majority do show significant increases in S/N, which are not seen in *any* of the follow-ups of search candidates, leading to the conclusion that no convincing detection candidates are present.

In the O1 run, where the run duration was $T_{\text{run}} = 1.12 \times 10^7$ s, the theoretical fractional loss of S/N will be

$$0.012 \left(\frac{T_{\text{drift}}}{10^6\ \text{s}}\right) \left(\frac{|\dot{f}|_{\text{drift}}}{10^{-12}\ \text{Hz s}^{-1}}\right)^2 \left(\frac{T_{\max}}{25000\ \text{s}}\right)^2. \tag{11}$$

Since our largest initial $T_{\max}$ value is 25,290 s, the impact on the initial search and the upper limit of spin wandering at or below this level would be negligible. Note that even spin wandering, which posed no complication for the initial search, could potentially be a limitation for the follow-up procedure, where $T_{\max}$ is increased by a factor of 4 at level 2 and a factor of 16 at level 3. In any event, the impact depends on the level of spin wandering present, which is still an area of open research.

[155] For comparison, the maximum spin wandering that could be tracked by the Viterbi analysis of Abbott et al. (2017f) is $|\dot{f}|_{\text{drift}} = (\Delta f_{\text{drift}})/T_{\text{drift}} = 1/(2T^2_{\text{drift}}) \approx 0.7 \times 10^{-12}$ Hz s$^{-1}$ at $T_{\text{drift}} = 10$ d $\approx 0.9 \times 10^6$ s.

## 5. Upper Limits

In the absence of a detection, we set upper limits on the strength of gravitational radiation from Sco X-1, as a function of frequency. We used as a detection statistic $\rho_{\max}$, the maximum statistic value observed in a 0.05 Hz band. We produced frequentist 95% upper limits via a combination of theoretical considerations and calibration with simulated signals, as explained in detail in Appendix A. The starting point was a Bayesian upper limit constructed using the expected statistical properties of the detection statistic and corrected for the reduction of sensitivity due to known lines. A series of simulated signal injections was then performed and used to estimate a global adjustment factor to estimate the amplitude at which a signal would have a 95% chance of increasing the $\rho_{\max}$ value in a band.

The procedure produced two sets of upper limits: a limit on $h_0$ including marginalization over the unknown inclination angle $\iota$, and an unmarginalized limit on the quantity $h_0^{\text{eff}}$ defined in Equation (2) to which the search is directly sensitive. The $h_0^{\text{eff}}$ upper limit can also be interpreted as a limit on $h_0$ subject to the assumption of circular polarization (optimal spin orientation corresponding to $\cos\iota = \pm 1$). It can be converted to a limit assuming linear polarization $\cos\iota = 0$ by multiplying by $\sqrt{8} = 2.83$. If we assume that the NS spin is aligned with the binary orbit (as one would expect for an NS spun up by accretion), $\iota \approx i \approx 44°$, we obtain a limit on $h_0$, which is the $h_0^{\text{eff}}$ upper limit multiplied by 1.35.

We show the marginalized and unmarginalized upper limits of this search in Figure 5, along with the other upper limits on Sco X-1 set with O1 data: the unmodeled stochastic radiometer (Ballmer 2006) results of Abbott et al. (2017d) and the directed search results of Abbott et al. (2017f) using Viterbi tracking of a hidden Markov model (Suvorova et al. 2016) to expand the applicability of the sideband search (Messenger & Woan 2007; Sammut et al. 2014; Aasi et al. 2015b) over the whole run. The present results improve on these by a factor of 3–4, yielding a marginalized limit of $h_0 \lesssim 2.3 \times 10^{-25}$ and an unmarginalized limit of $h_0^{\text{eff}} \lesssim 8.0 \times 10^{-26}$ at the most sensitive signal frequencies between around 100 Hz and 200 Hz. The marginalized 95% upper limits from Initial LIGO data (Abadie et al. 2011; Aasi et al. 2015b; Meadors et al. 2017) were all around $1.5 \times 10^{-25}$, so we have achieved an overall improvement of a factor of 6–7 from Initial LIGO to Advanced LIGO's first observing run, a combination of decreased detector noise and algorithmic improvements.

We also plot for comparison the torque balance level predicted by Equation (4). The marginalized limits on $h_0$ come closest to this level at 100 Hz, where they are within a factor of 3.4 of this theoretical level. In terms of $h_0^{\text{eff}}$, the torque balance level depends on the unknown value of the inclination $\iota$. For the most optimistic case of circular polarization ($\cos\iota = \pm 1$), our unmarginalized limit is a factor of 1.2 above the torque balance level near 100 Hz. Assuming linear polarization puts our limits within a factor of 3.5 of this level, and the most likely value of $\iota = 44$ corresponds to an upper limit curve a factor of 1.7 above the torque balance level, again near 100 Hz.

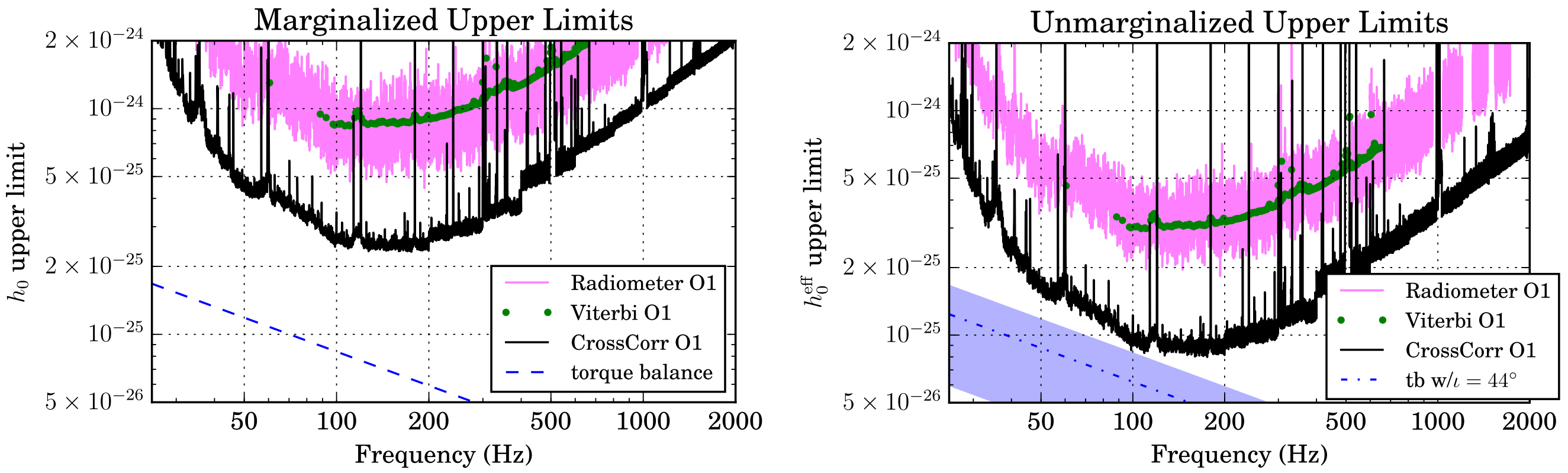

**Figure 5.** Upper limits from directed searches in O1 data. Left: upper limit on $h_0$, after marginalizing over the neutron star spin inclination $\iota$, assuming an isotropic prior. The dashed line shows the nominal expected level assuming torque balance (Equation (4)) as a function of frequency. Right: upper limit on $h_0^{\rm eff}$, defined in Equation (2). This is equivalent to the upper limit on $h_0$ assuming circular polarization. (Note that the marginalized upper limit on the left is dominated by linear polarization, and so is a factor of $\sim\sqrt{8}$ higher.) The shaded band shows the range of $h_0^{\rm eff}$ levels corresponding to the torque balance $h_0$ in the plot on the left, with circular polarization at the top and linear polarization on the bottom. The dotted–dashed line (labelled "tb w/$\iota = 44°$") corresponds to the assumption that the neutron star spin is aligned to the most likely orbital angular momentum, and $\iota \approx i \approx 44°$. (See Table 1.) For comparison with the "CrossCorr" results presented in this paper, we show the "unknown polarization" and "circular polarization" curves from the Viterbi analysis in Abbott et al. (2017f; dark green dots), as well as the 95% marginalized and circular polarization adapted from the Radiometer analysis in Abbott et al. (2017d; broad light magenta curve). Note that the Viterbi analysis reported upper limits for the 1 Hz bands, while the current CrossCorr analysis does so for the 0.05 Hz bands, and the Radiometer analysis for the 0.03125 Hz bands. This gives the upper limit curves for CrossCorr and especially Radiometer a "fuzziness" associated with noise fluctuations between adjacent frequencies rather than any physically meaningful distinction. When comparing 95% upper limits between the different analyses, it is therefore appropriate to look near the 95th percentile of this "fuzz" rather than at its bottom.

## 6. Outlook for Future Observations

We have presented the results of a search for GWs from Sco X-1 using data from Advanced LIGO's first observing run. The upper limits on the GW amplitude represent a significant improvement over the results from Initial LIGO and are within a factor of 1.2–3.5 of the benchmark set by the torque balance model, depending on assumptions about system orientation. Future observing runs (Abbott et al. 2016f) are expected to produce an improvement in the detector strain sensitivity of $\gtrsim$2.5. An additional enhancement will come with longer runs, as the amplitude sensitivity of the search scales as $T_{\rm obs}^{1/4}$. Algorithmic improvements that allow larger $T_{\rm max}$ with the same computing resources will also lead to improvements, as the sensitivity scales as $T_{\rm max}^{1/4}$ as well. A promising area for such an improvement is the use of resampling (Patel et al. 2010) to reduce the scaling of computing cost with $T_{\rm max}$ (G. D. Meadors et al. 2017, in preparation). (A similar method is used in the proposed semicoherent search described in Leaci & Prix 2015.) These anticipated instrumental and algorithmic improvements make it likely that search sensitivities will surpass the torque balance level over a range of frequencies (as projected in Whelan et al. 2015), and suggest the possibility of a detection during the advanced detector era, depending on details of the system such as GW frequency, inclination of the NS spin to the line of sight, and how close the system is to GW torque balance.

The authors gratefully acknowledge the support of the United States National Science Foundation (NSF) for the construction and operation of the LIGO Laboratory and Advanced LIGO as well as the Science and Technology Facilities Council (STFC) of the United Kingdom, the Max-Planck-Society (MPS), and the State of Niedersachsen/Germany for support of the construction of Advanced LIGO, and construction and operation of the GEO600 detector. Additional support for Advanced LIGO was provided by the Australian Research Council. The authors gratefully acknowledge the Italian Istituto Nazionale di Fisica Nucleare (INFN), the French Centre National de la Recherche Scientifique (CNRS), and the Foundation for Fundamental Research on Matter supported by the Netherlands Organisation for Scientific Research, for the construction and operation of the Virgo detector and the creation and support of the EGO consortium. The authors also gratefully acknowledge research support from these agencies as well as by the Council of Scientific and Industrial Research of India, Department of Science and Technology, India, Science & Engineering Research Board (SERB), India, Ministry of Human Resource Development, India, the Spanish Ministerio de Economía y Competitividad, the Vicepresidència i Conselleria d'Innovació Recerca i Turisme and the Conselleria d'Educació i Universitat del Govern de les Illes Balears, the National Science Centre of Poland, the European Commission, the Royal Society, the Scottish Funding Council, the Scottish Universities Physics Alliance, the Hungarian Scientific Research Fund (OTKA), the Lyon Institute of Origins (LIO), the National Research Foundation of Korea, Industry Canada and the Province of Ontario through the Ministry of Economic Development and Innovation, the Natural Science and Engineering Research Council Canada, Canadian Institute for Advanced Research, the Brazilian Ministry of Science, Technology, and Innovation, International Center for Theoretical Physics South American Institute for Fundamental Research (ICTP-SAIFR), Russian Foundation for Basic Research, the Leverhulme Trust, the Research Corporation, Ministry of Science and Technology (MOST), Taiwan, and the Kavli Foundation. The authors gratefully acknowledge the support of the NSF, STFC, MPS, INFN, CNRS, and the State of Niedersachsen/Germany for the provision of computational resources.

This paper has been assigned LIGO Document No. LIGO-P1600297-v24.

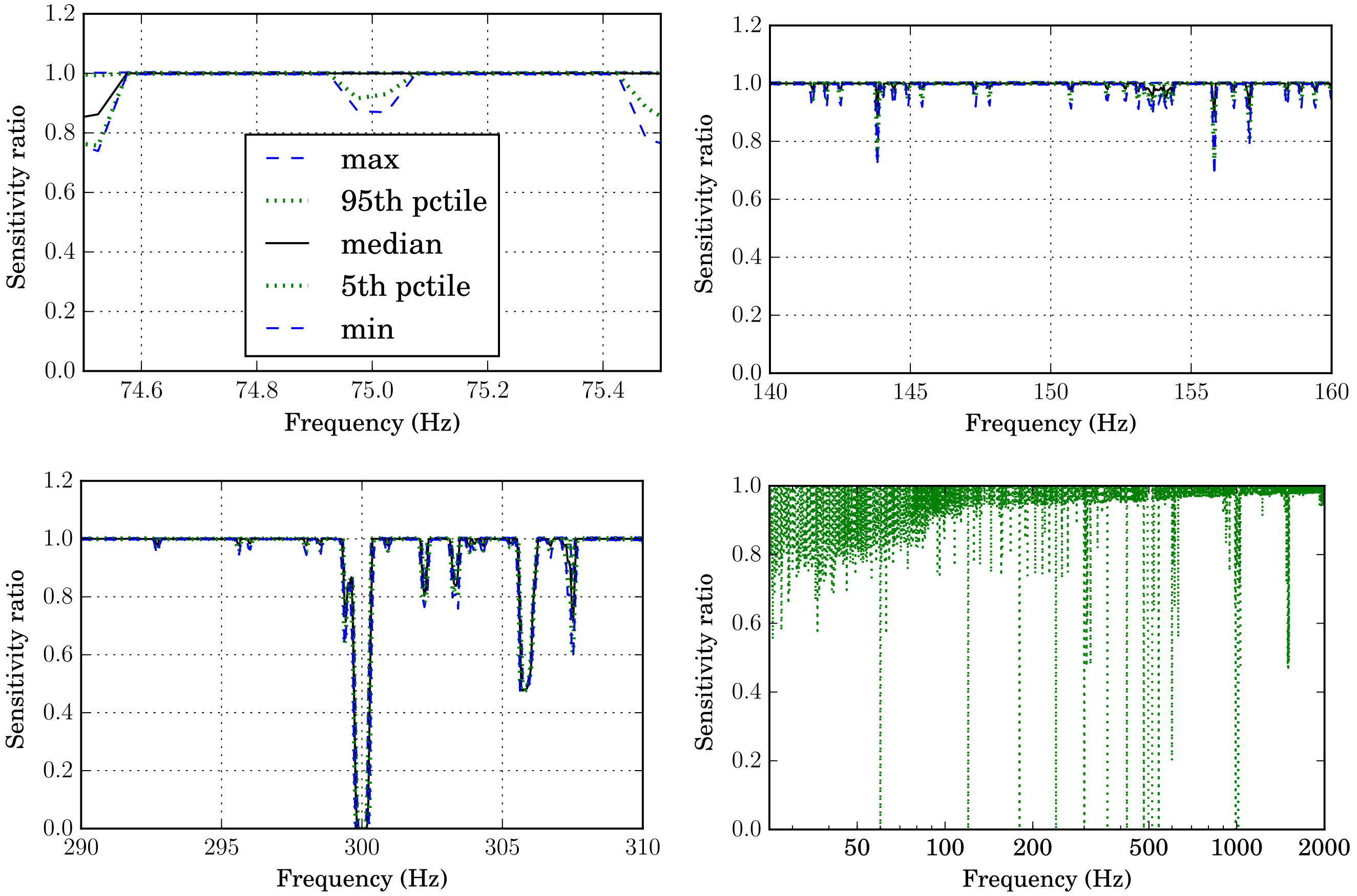

**Figure 6.** Impact of known lines on the sensitivity of the search. Fourier bins impacted by known lines are removed from the calculation of the statistic $\rho$ defined in Equation (5) and from the sensitivity $\vartheta = E[\rho]/(h_0^{\text{eff}})^2$ defined in Equation (7). For a given signal frequency $f_0$, data are removed at some times due to the time-varying Doppler shift which depends on the orbital parameter $a$ sin $i$. The effect is to lower $\vartheta$ relative to the value it would have if the lines were not removed; this "sensitivity ratio" goes to zero if all of the data relevant to a signal frequency $f_0$ are removed by the line. The first three plots contain illustrations of the percentiles of this ratio, taken over intervals of width 0.05 Hz. (There is a range of values in each frequency interval because of its finite width, and the range of $a$ sin $i$ values which determine the magnitude of the Doppler modulation.) Note that the broad line at 300 Hz (a harmonic of the 60 Hz AC power line) effectively nullifies the search at that frequency. The last plot shows the fifth percentile of the sensitivity ratio in 0.05 Hz intervals across the whole sensitivity band.

## Appendix A
## Details of the Upper Limit Method

The method used to set the upper limits for each 0.05 Hz band in Section 5 consisted of three steps:

1. An idealized 95% Bayesian upper limit was constructed using the posterior distribution pdf$(h_0^2|\rho_{\max})$ or pdf$([h_0^{\text{eff}}]^2|\rho_{\max})$.
2. A correction factor was applied in each 0.05 Hz band to account for the loss of sensitivity due to omission of data impacted by known lines.
3. A series of software injections was performed near the level of the 95% upper limit and used to empirically estimate a global correction factor for each upper limit curve based on the recovery or non-recovery of the injections.

### A.1. Idealized Bayesian Method

The Bayesian calculation assumes that all of the $\rho$ values for templates in the initial search represent independent Gaussian random variables with unit variance; one has mean $[h_0^{\text{eff}}]^2\vartheta$ and the others have zero mean. Note that different regions of the orbital parameter space have different coherence times $T_{\max}$ and therefore $\vartheta$ values (cf. Equation (7)). The method produces a sampling distribution pdf$(\rho_{\max}|[h_0^{\text{eff}}]^2)$, marginalizing over the location of the signal in orbital parameter space.

This sampling distribution is used to construct a posterior distribution pdf$([h_0^{\text{eff}}]^2|\rho_{\max})$ assuming a uniform prior in $(h_0^{\text{eff}})^2$, and this is used to produce a 95% Bayesian upper limit on $(h_0^{\text{eff}})^2$ according to

$$\int_0^{(h_0^{\text{eff}})^2_{95\%}} d[h_0^{\text{eff}}]^2 \, \text{pdf}([h_0^{\text{eff}}]^2|\rho_{\max}) = 0.95. \qquad (12)$$

To produce an upper limit on the intrinsic strength $h_0$, we assume a prior that is uniform in $h_0^2$ and $\cos\iota$, repeat the calculation above, and numerically marginalize over $\cos\iota$ to obtain a posterior pdf$(h_0^2|\rho_{\max})$.

### A.2. Correction for Known Lines

Although we calculate a single $\vartheta$ value for each of the 18 search regions for a given 0.05 Hz band and use it in the calculation, the search can in principle have a different $\vartheta$ value for each template. This is because of the correction which omits data contaminated by Doppler-modulated known instrumental lines from the sum in Equation (7), a process that depends on the signal frequency $f_0$ as well as the projected orbital semimajor axis $a \sin i$. In each 0.05 Hz band, we estimate the distribution of the ratio of the actual $\vartheta$ to the band-wide $\vartheta$; the percentiles of this distribution are illustrated in Figure 6. We divide by the fifth percentile of this distribution (shown in the last panel of Figure 6) to produce corrected $h_0^2$ and $(h_0^{\rm eff})^2$ upper limits.

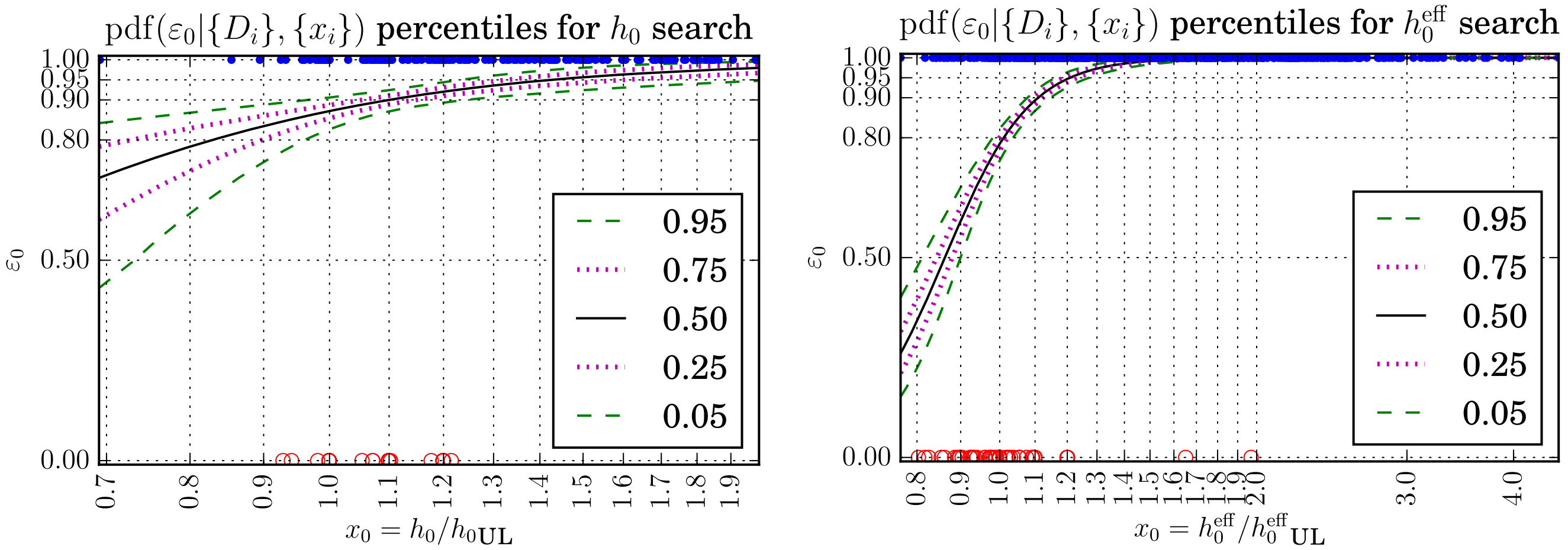

**Figure 7.** Estimation of efficiency from recovery of simulated signals injected in the software. At left, the results of the 376 injections with amplitude $h_0$ specified in terms of the uncorrected marginalized upper limit $h_{0\rm UL}$ are shown as black dots, with recovered injections (those that increased the maximum S/N $\rho_{\rm max}$ in the relevant 0.05 Hz band) shown as blue dots on the $\varepsilon_0 = 1$ line and unrecovered injections shown as red circles at $\varepsilon_0 = 0$. The recovered and unrecovered injections are used to produce a posterior pdf($\alpha$, $\beta|\{D_i\}$, $\{x_i\}$) according to Equation (13), and this is used to generate posterior distributions pdf($\varepsilon_0|\{D_i\}$, $\{x_i\}$) at a range of signal strengths $x_0 = h_0/h_{0\rm UL}$ according to Equation (14); the 5th, 25th, 50th, 75th, and 95th percentiles are calculated from these distributions as a function of $x_0$. The $x_0$ value at which the posterior expectation $E(\varepsilon_0|\{D_i\}, \{x_i\})$ crosses 0.95 is used as a correction factor by which we multiply $h_{0\rm UL}$ to produce the final marginalized upper limit shown in Figure 5. At right, we do the same thing for $h_0^{\rm eff}$, using the full set of 754, and derive a correction factor by which to multiply the unmarginalized upper limit $h_{0\rm UL}^{\rm eff}$. Note that in the $h_0$ search the value of $x_0$ corresponding to $\epsilon_0 = 0.95$ is less accurately determined than that for the $h_0^{\rm eff}$ search, both because of the smaller number of applicable injections and because the detection efficiency depends more weakly on $h_0$.

### *A.3. Empirical Adjustment from Software Injections*

We performed a series of re-analyses of the data with a total of 754 simulated signals ("software injections") added to the data stream to validate the upper limits including the known line correction. The signals were generated over signal frequencies from 25 to 500 Hz, some with $h_0$ set to some multiple of the marginalized 95% upper limit $h_{0\rm UL}$, and others with $h_0^{\rm eff}$ set to some multiple of the unmarginalized 95% upper limit $h_{0\rm UL}^{\rm eff}$. We defined "recovery" of the injection as an increase in the maximum detection statistic $\rho_{\rm max}$ compared to the results with no signal present. (Follow-up of the recovered injections that crossed the relevant $\rho$ threshold was also performed as a way of testing our follow-up procedure, as described in Section 4.) We find that the fraction of signals of each type recovered when the injection is done at the upper limit level to be slightly below the expected 95%.[156] This is to be expected, as there are various approximations in the method, such as the tolerated mismatch in the initial parameter space grid and the acceptable loss of S/N due to finite-length SFTs, which should lead to an S/N slightly less than that predicted by Equation (6).

To estimate empirically the amount by which the upper limits should be scaled to produce a 95% injection recovery efficiency, we apply the method described in Whelan (2015) and used to produce the efficiency curves in Messenger et al. (2015). We posit a simple sigmoid model where the efficiency of the search as a function of signal strength $x$ is assumed to be $\varepsilon(x;\, \alpha,\, \beta) = (1 + e^{-\alpha(\ln x - \beta)})^{-1}$ and construct the posterior from the recovery data ($D_i = 1$ if the signal $i$ was recovered, 0 if not):

$$\mathrm{pdf}(\alpha,\, \beta|\{D_i\},\, \{x_i\}) \propto \prod_i \varepsilon(x_i;\, \alpha,\, \beta)^{D_i} \times (1 - \varepsilon(x_i;\, \alpha,\, \beta))^{1-D_i}\mathrm{pdf}(\alpha,\, \beta). \quad (13)$$

With sufficient data, the prior should be irrelevant, but we take a noninformative prior pdf($\alpha$, $\beta$) $\propto \alpha^{-1}$ and define the signal strength $x$ as the $h_0$ or $h_0^{\rm eff}$ of the injection divided by the corresponding upper limit. We can then construct, at any signal level $x_0$, the posterior on the efficiency $\varepsilon_0 = \varepsilon(x_0;\, \alpha,\, \beta)$, marginalized over $\alpha$ and $\beta$:

$$\mathrm{pdf}(\varepsilon_0|\{D_i\},\, \{x_i\}) = \int_0^\infty d\alpha \int_{-\infty}^\infty d\beta\ \mathrm{pdf}(\alpha,\, \beta|\{D_i\}, \times \{x_i\})\ \delta(\varepsilon_0 - \varepsilon(x_0;\, \alpha,\, \beta)). \quad (14)$$

The posterior distributions of efficiency are shown in Figure 7. We define the correction factor to be the $x_0$ at which the expectation value $\int_0^1 d\varepsilon_0\ \mathrm{pdf}(\varepsilon_0|\{D_i\},\, \{x_i\})$ crosses 95%.

A total of eight sets of injections were performed, four with $h_0$ at a specified multiple of $h_{0\rm UL}$, and four with $h_0^{\rm eff}$ at a specified multiple of $h_{0\rm UL}^{\rm eff}$. The multipliers were 1.0, 1.1, 1.2, and a random value between 1.1 and 2.0 chosen from a log-uniform distribution. For the unmarginalized $h_0^{\rm eff}$ upper limit, we use all eight sets of injections, 754 in total and find the expectation value of the efficiency crosses 95% at $h_0^{\rm eff}/h_{0\rm UL}^{\rm eff} \approx 1.21$. This factor has been applied to $h_{0\rm UL}^{\rm eff}$ to produce the upper limits in Figure 5.

For the marginalized $h_0$ upper limit, we must restrict ourselves to the four injection sets which specified $h_0/h_{0\rm UL}$. This is because our search is primarily sensitive to $h_0^{\rm eff}$, and specifying $h_0^{\rm eff}$ while choosing the inclination angle $\iota$ randomly implies anticorrelations between $h_0$ and $|\cos\iota|$. Signals with high $h_0$ values will tend to be those with unfavorable polarization, and therefore not be any easier to detect. Using the 376 applicable injections, we estimate the 95% efficiency at $h_0/h_{0\mathrm{UL}} \approx 1.44$ and use this factor when generating the final upper limit shown in Figure 5. Note that this is less well determined than the factor for the unmarginalized $h_0^{\mathrm{eff}}$ upper limit. This is both because of the smaller number of injections used and because $h_0$ correlates less well with detectability than $h_0^{\mathrm{eff}}$. However, the upper limit curve for $h_0$ is very close to the unmarginalized upper limit assuming linear polarization ($\cos \iota = 0$), which is consistent with the expectation that the 95% upper limit will be dominated by this worse-case scenario.

[156] The fraction of signals recovered is a frequentist statement, as opposed to the Bayesian upper limit constructed from the posterior, but the two types of upper limits are related closely enough (see, for example, Rover et al. 2011) that the fraction should be close to 95%.

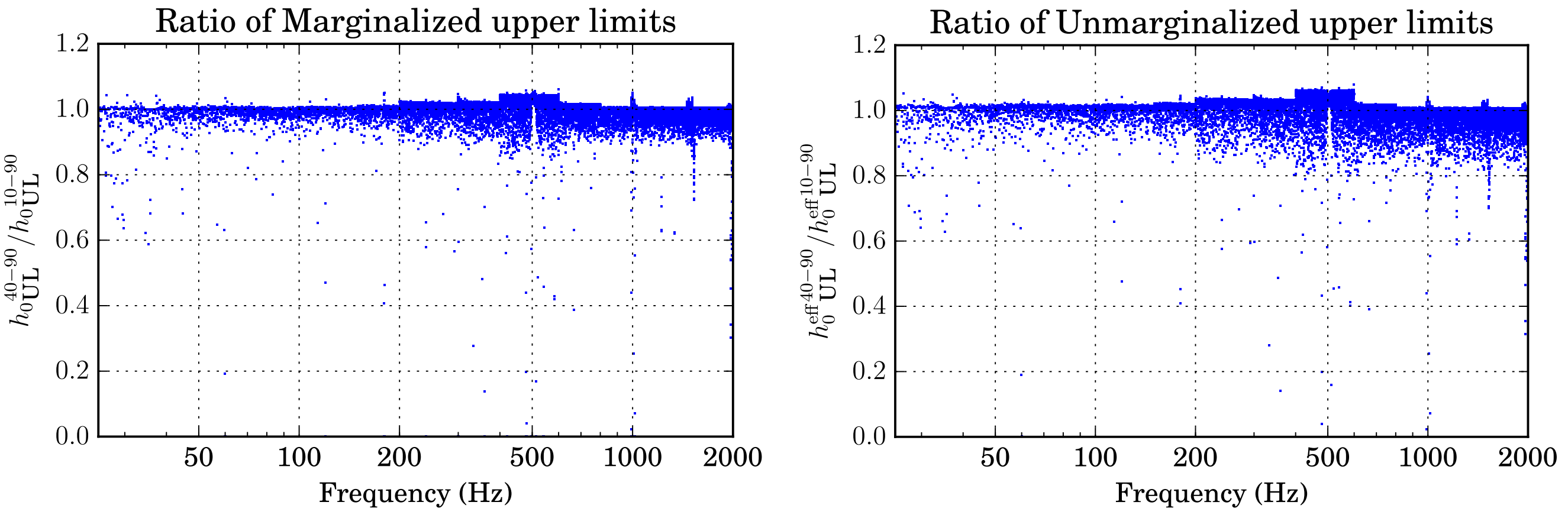

**Figure 8.** Comparison of upper limits constructed by restricting attention to $a \sin i \geqslant 1.44$ lt-s ($K_1 \geqslant 40$ km s$^{-1}$) to those from the original search. The results are generally comparable; we plot the ratio of the upper limits rather than reproducing the curves in Figure 5, because the changes in the latter would barely be noticeable. The step-like features that are visible are due to the details of the search (such as $T_{\max}$ values) being different in different frequency ranges listed in Table 3.

## Appendix B
## Results with a Constrained Semimajor Axis

As noted in Table 2, the range of $a \sin i$ values searched was chosen based on preliminary information from Wang (2017), which constrained the projected orbital velocity $K_1$ to lie between 10 and 90 km s$^{-1}$. This was subsequently refined to between 40 and 90 km s$^{-1}$. For comparison, we recomputed the upper limits, discarding the results of searches with $a \sin i \leqslant 1.44$ lt-s, corresponding to the nine bottom-most parameter space cells shown in Figure 2. The results were not significantly different (for instance, they were barely noticeable on plots like Figure 5), but for illustration we plot in Figure 8 the ratio of the two sets of upper limits. A bigger impact of the refined parameter space will be seen in future runs, when computing resources can be concentrated on the allowed range of $a \sin i$ values.

## References

Aasi, J., Abadie, J., Abbott, B. P., et al. 2014a, ApJ, 785, 119
Aasi, J., Abbott, B. P., Abbott, R., et al. 2014b, PhRvD, 90, 062010
Aasi, J., Abbott, B. P., Abbott, R., et al. 2015a, ApJ, 813, 39
Aasi, J., Abbott, B. P., Abbott, R., et al. 2015b, PhRvD, 91, 062008
Aasi, J., Abbott, B. P., Abbott, R., et al. 2015c, CQGra, 32, 074001
Aasi, J., Abbott, B. P., Abbott, R., et al. 2016, PhRvD, 93, 042007
Abadie, J., Abbott, B. P., Abbott, R., et al. 2010, ApJ, 722, 1504
Abadie, J., Abbott, B. P., Abbott, R., et al. 2011, PhRvL, 107, 271102
Abbott, B., Abbott, R., Adhikari, R., et al. 2007a, PhRvD, 76, 082001
Abbott, B., Abbott, R., Adhikari, R., et al. 2007b, PhRvD, 76, 082003
Abbott, B. P., Abbott, R., Abbott, T. D., et al. 2016a, PhRvD, 94, 042002
Abbott, B. P., Abbott, R., Abbott, T. D., et al. 2016b, PhRvX, 6, 041015
Abbott, B. P., Abbott, R., Abbott, T. D., et al. 2016c, PhRvL, 116, 061102
Abbott, B. P., Abbott, R., Abbott, T. D., et al. 2016d, PhRvL, 116, 241103
Abbott, B. P., Abbott, R., Abbott, T. D., et al. 2016e, ApJL, 832, L21
Abbott, B. P., Abbott, R., Abbott, T. D., et al. 2016f, LRR, 19, 1
Abbott, B. P., Abbott, R., Abbott, T. D., et al. 2017a, PhRvD, 95, 042003
Abbott, B. P., Abbott, R., Abbott, T. D., et al. 2017b, ApJ, 841, 89
Abbott, B. P., Abbott, R., Abbott, T. D., et al. 2017c, PhRvL, 118, 121101
Abbott, B. P., Abbott, R., Abbott, T. D., et al. 2017d, PhRvL, 118, 121102
Abbott, B. P., Abbott, R., Abbott, T. D., et al. 2017e, ApJ, 839, 12
Abbott, B. P., Abbott, R., Abbott, T. D., et al. 2017f, PhRvD, 95, 122003
Ballmer, S. W. 2006, CQGra, 23, S179
Bildsten, L. 1998, ApJL, 501, L89
Bradshaw, C. F., Fomalont, E. B., & Geldzahler, B. J. 1999, ApJL, 512 L121
Dhurandhar, S., Krishnan, B., Mukhopadhyay, H., & Whelan, J. T. 2008, PhRvD, 77, 082001
Fomalont, E. B., Geldzahler, B. J., & Bradshaw, C. F. 2001, ApJ, 558, 283
Galloway, D. K., Premachandra, S., Steeghs, D., et al. 2014, ApJ, 781, 14
Goetz, E., & Riles, K. 2011, CQGra, 28, 215006
Haskell, B., Andersson, N., D'Angelo, C., et al. 2015a, in Astrophysics and Space Science Proc. 40, Gravitational Wave Astrophysics, ed. C. F. Sopuerta (Switzerland: Springer), 85
Haskell, B., Priymak, M., Patruno, A., et al. 2015b, MNRAS, 450, 2393
Jaranowski, P., Krolak, A., & Schutz, B. F. 1998, PhRvD, 58, 063001
Leaci, P., & Prix, R. 2015, PhRvD, 91, 102003
Meadors, G. D., Goetz, E., & Riles, K. 2016, CQGra, 33, 105017
Meadors, G. D., Goetz, E., Riles, K., Creighton, T., & Robinet, F. 2017, PhRvD, 95, 042005
Messenger, C. 2011, PhRvD, 84, 083003
Messenger, C., Bulten, H., Crowder, S., et al. 2015, PhRvD, 92, 023006
Messenger, C., & Woan, G. 2007, CQGra, 24, S469
Papaloizou, J., & Pringle, J. E. 1978, MNRAS, 184, 501
Patel, P., Siemens, X., Dupuis, R., & Betzwieser, J. 2010, PhRvD, 81, 084032
Rover, C., Messenger, C., & Prix, R. 2011, arXiv:1103.2987
Sammut, L., Messenger, C., Melatos, A., & Owen, B. 2014, PhRvD, 89, 043001
Steeghs, D., & Casares, J. 2002, ApJ, 568, 273
Suvorova, S., Sun, L., Melatos, A., Moran, W., & Evans, R. 2016, PhRvD, 93, 123009
Wagoner, R. V. 1984, ApJ, 278, 345
Wang, L. 2017, PhD thesis, Warwick Univ.
Watts, A., Krishnan, B., Bildsten, L., & Schutz, B. F. 2008, MNRAS, 389, 839
Wette, K., Owen, B. J., Allen, B., et al. 2008, CQGra, 25, 235011
Whelan, J. T., Sundaresan, S., Zhang, Y., & Peiris, P. 2015, PhRvD, 91, 102005
Whelan, J. T. 2015, Bayesian Estimation of Parametrized Efficiency, LIGO Graphical Presentation LIGO-G1500977, https://dcc.ligo.org/LIGO-G1500977/public